\documentclass[preprint,12pt]{elsarticle}

\usepackage{amsthm, amsmath, amsfonts, amssymb, bbm, bm}
\usepackage{blindtext}
\usepackage[colorlinks=true,
            linkcolor=blue,
            urlcolor=blue]{hyperref}

\DeclareMathOperator*{\argmin}{arg\,min}

\theoremstyle{plain}
\newtheorem{Proposition}{Proposition}

\journal{Reliability Engineering and System Safety}

\begin{document}

\begin{frontmatter}

\title{Fusion of heterogeneous data for robust degradation prognostics}

\author[EDF,ENS,LISN]{Edgar Jaber}
\author[EDF]{Emmanuel Remy}
\author[EDF]{Vincent Chabridon}
\author[ENS]{Mathilde Mougeot}
\author[LISN]{Didier Lucor}
\affiliation[EDF]{organization={EDF R\&D},
            addressline={6 Quai Watier},
            city={Chatou},
            postcode={78401},
            country={France}}
\affiliation[ENS]{organization={Université Paris-Saclay, CNRS, ENS Paris-Saclay, Centre Borelli},
            city={Gif-sur-Yvette},
            postcode={91190},
            country={France}}
\affiliation[LISN]{organization={Université Paris-Saclay, CNRS, Laboratoire Interdisciplinaire des Sciences du Numérique},
            city={Orsay},
            postcode={91405},
            country={France}}

\begin{abstract}
\emph{Assessing the degradation state of an industrial asset first requires evaluating its current condition and then projecting the forecast model trajectory to a predefined prognostic threshold, thereby estimating its remaining useful life (RUL). Depending on the available information, two primary categories of forecasting models may be used: model-based simulation codes and data-driven (machine learning) approaches. Combining both modelling approaches may enhance prediction robustness, especially with respect to their individual uncertainties. This paper introduces a methodology for fusion of heterogeneous data in degradation prognostics. The proposed modular approach acts iteratively on a computer model's uncertain input variables by combining kernel-based sensitivity analysis for variable ranking with a Bayesian framework to inform the priors with the heterogeneous data - and adds a Kalman based smoothing step for reducing uncertainties on the prognostics horizon. Additionally, we propose an integration of an aggregate surrogate modeling strategy for computationally expensive degradation simulation codes. The methodology updates the knowledge of the computer code input probabilistic model and reduces the output uncertainty. As an application, we illustrate this methodology on a toy model from crack propagation based on Paris law as well as a complex industrial clogging simulation model for nuclear power plant steam generators, where data is intermittently available over time.}
\end{abstract}

\begin{highlights}
\item This paper provides a full methodology for fusion of heterogeneous data for a computationally intensive degradation simulation model. 
\item The proposed modular approach performs offline data assimilation in two steps. Firstly, it employs a Bayesian model updating (BMU) step using kernel sensitivity analysis techniques to rank and evaluate the time-varying importance of input variables and then provides a specific Bayesian technique to sample from the data-informed posterior distributions using Monte Carlo Markov chain (MCMC) techniques. Secondly it makes use of Ensemble Kalman smoothing methods for full state updating and subsequent uncertainty reduction.
\item The proposed BMU method overcomes the curse of dimensionality for high-dimensional MCMC posterior estimation by iteratively updating the marginal distributions of individual influent input variables assuming independent marginals and by measuring the data-informed posterior compared to the prior by computing the Kullback-Leibler divergence.
\item The proposed method allows to make robust probabilistic predictions of the remaining useful life of an asset by propagating data-informed posteriors of the input variables in the computer simulation model and comparing it to non-informed prior distribution.
\item The proposed method is well suited for expensive-to-evaluate simulation models by integrating a surrogate modeling step and proposes a way to integrate the induced metamodeling bias with a Monte Carlo aggregation approach. 
\item Benefits of our method are showcased first on a controlled application to Paris-Erdogan's law for crack growth propagation with a fictitious material and then on a real digital twin for steam generators clogging in nuclear power plants.
\end{highlights}

\begin{keyword}
Data fusion \sep Prognostics and Health Management \sep Heterogeneous data \sep Surrogate models \sep Global sensitivity analysis \sep Bayesian model updating \sep Monte Carlo Markov Chains \sep Ensemble Kalman smoothing \sep Nuclear power plants \sep Digital twins
\end{keyword}

\end{frontmatter}

\section{Introduction}
\label{sec1}

Prognostics and Health Management (PHM) \citep{biggio_prognostics_2020} is an interdisciplinary engineering domain that unifies multiple fields into a structured framework to facilitate the maintenance of industrial assets. Prognostics entails assessing the current condition of an asset (called the diagnostics step) and projecting its degradation trajectory to a predefined threshold (called the prognostics step), thereby estimating the \emph{Remaining Useful Life} (RUL) \citep{escobet_fault_2019} of the asset. The RUL is defined as the time interval between the current moment and the estimated future operating time at which the predicted degradation indicator reaches a specified threshold, signifying the necessity for maintenance of the asset while possibly ensuring a safety margin before reaching a critical failure. Mathematically, denoting a positive, scalar time-dependent degradation indicator of a system as $(t\mapsto \delta(t))$, a predefined threshold as $D\in\mathbb{R}_{+}$, and the current time as $t_{P}$, the RUL can be formulated as:
\begin{equation}
    \text{RUL}(D) = \argmin_{t>t_{P}}\{\delta(t) \geq D\}.
\end{equation}
In the literature, two primary categories of prognostics models exist: model-based and data-driven approaches \citep{escobet_fault_2019}. The former relies on physics of failure principles and engineering models to describe degradation dynamics. It has demonstrated historically a strong efficiency in diverse applications, such as monitoring pneumatic valves \citep{daigle_2011}, lithium-ion batteries \citep{zhang_2011}, and nuclear heat removal subsystems \citep{liu_2017}. 

Conversely, data-driven models leverage statistical and machine learning techniques trained on measured operational data. Some examples of these methods include convolutional neural networks (CNNs) \citep{li_2018} and long short-term memory (LSTM) recurrent neural networks \citep{shi_2021} for predicting the RUL of turbofan engines, as well as support vector machines \citep{garcianieto_2015} and adjacent difference neural networks \citep{zhao_2017} for estimating the RUL of aircraft engines, among others \citep{zio_prognostics_2022}.

When separate forecasting approaches exhibit some limitations such as high uncertainty in the simulation model output or lack of generalization due to small training data set, hybrid strategies such as combining model-based and/or data-driven models, may be employed to enhance prediction robustness \citep{zio_prognostics_2022}. This combination is particularly useful when the physical model lacks precision in capturing system complexity or when the data-driven approach is insufficiently robust to account for the underlying physics. Among hybrid prognostic strategies, data assimilation techniques are widely adopted, yielding robust RUL estimates in applications such as crack growth modeling \citep{orchard_2007, orchard_2009, zio_2011, baraldi_2013}, lithium-ion battery degradation \citep{dong_2014, celaya_2012, he_2011}, and water tank systems \citep{chen_2012}. A common approach within data assimilation uses filtering techniques, including Kalman filters \citep{kalman_1960} and their various extensions, but also particle filters \citep{jouin_filters_review_2016} for nonlinear state models. 

Recent literature has also made important advances in these hybrid directions. Physics-informed Gaussian Process models, for instance, incorporate priors derived from physics simulations and have shown success in fatigue crack growth prediction \citep{jiang2025physicsinformed}. Deep learning-based frameworks, such as temporal domain adaptation networks \citep{xu2024temporal}, have been proposed to address operational shifts in data distributions. Semi-stochastic filtering techniques inspired by Wiener processes have also been introduced to infer RUL from indirect condition monitoring data \citep{xiaosheng2024wiener}, and hybrid models combining deep neural networks and statistical filters have improved robustness in lithium-ion battery applications \citep{duan2025interactive}. As for applications to power generation, systematic reviews such as the recent \citep{cheng2025diagnostics} highlight that although diagnostics have matured,  prognostic capabilities remain underdeveloped, especially for safety-critical and complex systems like nuclear power plants. Moreover, as emphasized in \citep{cuesta2025review}, practical applications of PHM still face major barriers including data sparsity, heterogeneity of sources, and limited physical observability. Despite these advances, several gaps remain:

\begin{itemize}
    \item Most methods assume access to dense, regularly sampled degradation trajectories, which rarely exist in real industrial environments.
    \item Few works address how to handle \textit{heterogeneous data sources}— such as simulation outputs, sparse inspections, or indirect proxy variables — in a unified probabilistic framework.
    \item Although surrogate models are often used to reduce computational costs, their aggregation and uncertainty contribution are typically treated as external steps rather than being integrated into the calibration process.
    \item Input relevance and dimensionality reduction are often handled heuristically or ignored altogether, limiting interpretability and scalability.
\end{itemize}

This paper addresses these shortcomings by introducing a principled and scalable methodology for RUL prognostics based on simulation-assisted Bayesian inference, with explicit fusion of sparse and heterogeneous data. Our contributions are summarized as follows:

\begin{enumerate}
    \item We propose a full data fusion procedure including full state input variables and output probabilistic update by conditioning  to heterogeneous groups of data. It includes a novel iterative Bayesian model updating (BMU) framework as well as a Kalman-based trajectories smoothing step. This allows to perform robust offline non-parametric probabilistic degradation prognostics.
    
    \item The BMU is guided by global sensitivity analysis using the Hilbert-Schmidt Independence Criterion (HSIC). We identify and update only the most influential input variables. This departs from prior work such as \citep{jiang2025physicsinformed}, which uses physics-based priors but does not perform selective updating or rank inputs.

    \item We develop a multi-source data fusion scheme, where each data type (e.g., simulations, inspections, monitoring signals) is modeled with a custom likelihood and uncertainty structure. This extends the single-source assumptions found in \citep{liu2024boxcox}, \citep{xu2024temporal}, and \citep{basora2025benchmark}, where training relies on clean or fully observed degradation trajectories.

    \item We incorporate surrogate model aggregation into the inference process for expensive computer codes, using Dirichlet-based weighting and explicit modeling of aggregation bias. While surrogates are widely used (e.g., \citep{duan2025interactive}, \citep{cuesta2025review}), their uncertainty and contribution to posterior variance are rarely integrated within the core Bayesian loop.
    
    \item We apply our methodology to a toy model illustrating crack propagation based on Paris-Erdogan's law and to a realistic digital twin use case in power generation, namely clogging prognostics in nuclear steam generators. In the latter, simulation-based models and sparse field data are reconciled for assisting operational maintenance planning. Unlike studies such as \citep{cheng2025diagnostics}, which survey methods in the nuclear sector but don't propose integration strategies, our implementation offers a deployable solution for such digital twins.
\end{enumerate}

The rest of this paper is structured as follows: section \ref{sec2} introduces the notations used throughout the paper as well as the overall methodology for robust fusion of heterogeneous data in degradation prognostics. Section \ref{sec3} presents the mathematical background of the methods used as well as details on the different algorithms. Section \ref{sec4} details the application of the methodology first on the crack propagation model and then the simulation model for clogging of steam generators in nuclear power plants. Section \ref{sec5} discusses the results and work perspective before concluding the paper.

\section{Notations and overall presentation of the methodology}
\label{sec2}
\subsection{Main notations}
\label{sec21}
\noindent Throughout this paper $g :\mathcal{X}\to \mathbb{R}^{N}$ is a measurable function representing time-varying simulation model with input variables $\bm{X}\in\mathcal{X}\subset\mathbb{R}^{d}$ equipped with their distribution $\mu_{\bm{X}}$. The pushforward operator is denoted $g\#$, defined on the space of probability distributions on $\mathcal{X}$ and with values in the space of probability distributions on $\mathbb{R}^{N}$, represents the image-measure through function $g$. The projector on the $k$-th coordinate of a vector $v\in\mathbb{R}^{N}$ is denoted by $\text{pr}_{k}\circ v = v_{k}$. The indicator function of a measurable set $A$ is denoted by $\bm{1}_{A}$. Probability densities is denoted by $p(.)$ and the corresponding conditioning on random variable $(\bm{Z} = \bm{z})$ is denoted by $p(.|\bm{Z}) = p(.|\bm{Z} = \bm{z})$. The expected value with respect to a measure $\mu$ is denoted by $\mathbb{E}_{\mu}[.]$, or alternatively for $X\sim \mu_{X}$, $\mathbb{E}_{X}[.]$. $\lVert . \rVert_{p}$ will correspond to the Euclidian norm on $\mathbb{R}^{p}$. A statistical surrogate model of the simulation code is denoted by $\widehat{g}$. A $n$-sample of $\bm{X} = (X_{1},\ldots,X_{d})$ is denoted by $\{\bm{X}^{(i)}\}_{i=1}^{n} = \{(X^{(i)}_{1},\ldots,X^{(i)}_{d})\}_{i=1}^{n}$. The $\ell^1$-sphere on the unit hypercube $[0,1]^{p}$ is the simplex $\Delta^{p-1} = \{\bm{w} \in[0,1]^p,\;w_1 + \ldots + w_p = 1\}$. The normal distribution with mean $m$ and covariance $K$ is written $\mathcal{N}(m,K)$. The Gamma distribution is denoted by $\mathcal{G}(\alpha, \beta)$, with probability density function $f_{\alpha,\beta}(x) = \frac{1}{\Gamma(\alpha)}x^{\alpha-1} \beta^{\alpha}\exp(-\beta x)$. The Dirichlet distribution $\text{Dir}^{p-1}$ corresponding to a uniform distribution on the simplex $\Delta^{p-1}$ is denoted by $\text{Dir}^{p-1} = \text{Dir}(1_{p})$ where $1_p = (1,\ldots,1)\in\mathbb{R}^p$, its density function is given by $f(\bm{w}) = (p-1)!\;\bm{1}_{\Delta^{p-1}}(\bm{w})$. When necessary, other notations will be introduced all along the paper for the sake of conciseness.

\subsection{General presentation of the methodology}
\label{subsec22}
\begin{figure}[h!]
    \centering
    \includegraphics[width=1.0\textwidth]{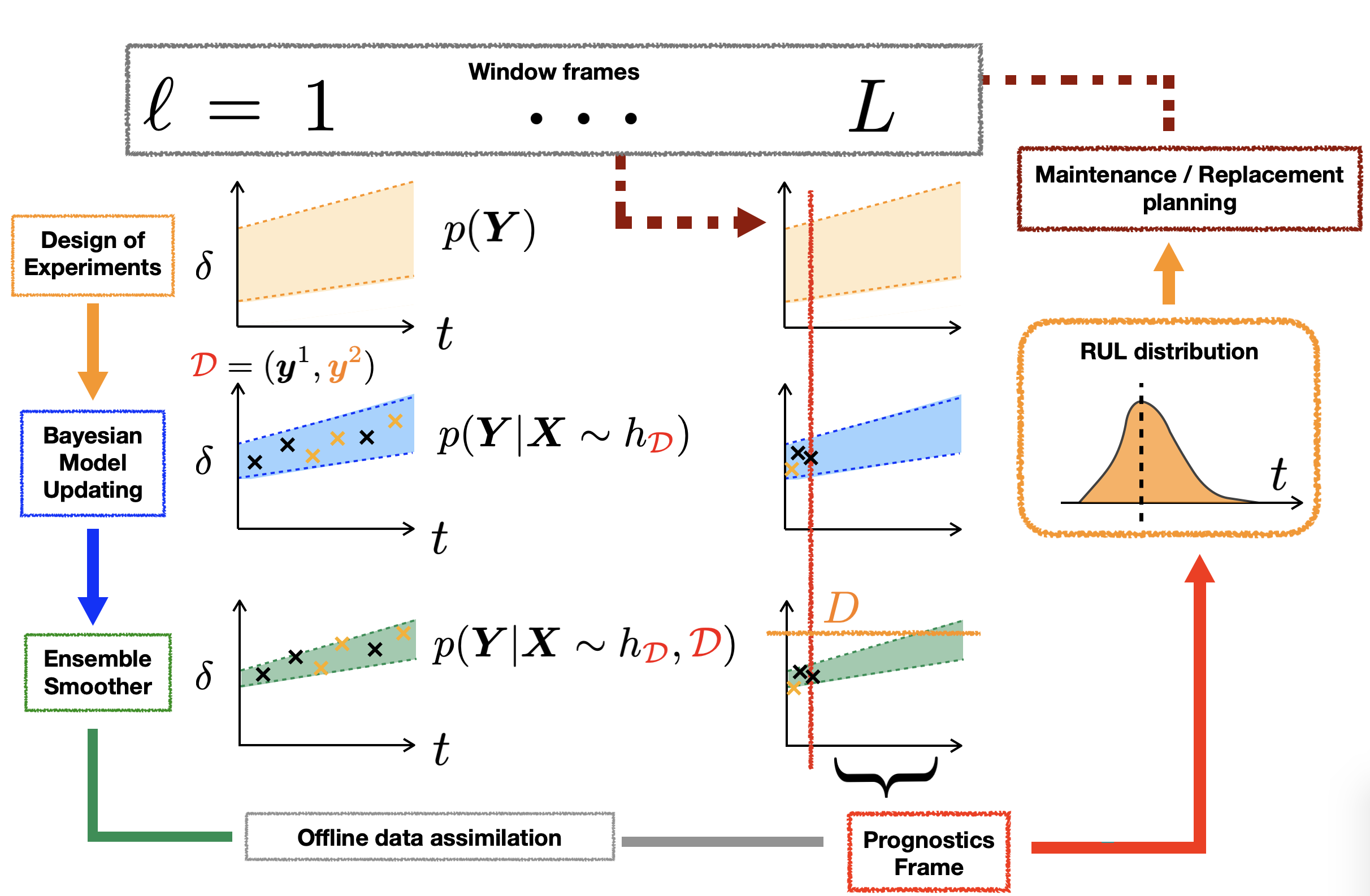}
    \caption{Proposed methodology for the offline data assimilation}
    \label{fig:offline_hybrid}
\end{figure}

In our context, real-time data is not available, making standard filtering approaches impractical; instead, we focus on assimilating all available data in a post hoc manner. The simulation model $g:\bm{\mathcal{X}}\subset\mathbb{R}^{d} \to \mathbb{R}^{N}$, with uncertain inputs $\bm{X} = (X_{1},\ldots,X_{d}) \sim \mu_{\bm{X}}$, produces a full trajectory for each input sample:
\begin{equation}
    g(\bm{x}_{0}) = (g(t_{1},\bm{x}_{0}), \ldots, g(t_{N},\bm{x}_{0})) \in \mathbb{R}^{N},
\end{equation}
where $\text{pr}_{\ell}\circ g(\bm{X})$ or $g(t_{\ell},\bm{X})$ represents the degradation index $\delta(t_{\ell})$. The model is treated as a grey-box: the underlying physics are known, but the code itself cannot be modified. We also consider $q$ heterogeneous degradation data groups, originating from different sensors or statistical models with varying fidelities. These are collected as $\mathcal{D} = (\bm{y}^{1},\ldots,\bm{y}^{q})$, with each $\bm{y}^{i}\in\mathbb{R}^{m_{i}}$ corresponding to distinct time indices $\mathcal{J}_{i}$, such that $\mathcal{J} = \cup_{i=1}^{q}\mathcal{J}_{i}$. For each group and time index:
\begin{equation}
    y^{i}(t_{\ell}) = \delta(t_{\ell}) + \eta^{i}_{\ell},
\end{equation}
where $\eta^{i}_{\ell} \sim \mathcal{N}(0,\sigma_{i}^{2})$ and $\sigma_{i}^{2}$ is the known noise variance for group $i$. Sequential data assimilation is performed over predefined $L$ time windows within the output time discretization. Figure~\ref{fig:offline_hybrid} summarises the proposed methodology. For each window frame $\ell=1,\ldots,L$, start with a design of experiments from the original computer model, apply the BMU methodology and sample from the pushforward of the updated inputs. On this ensemble, apply a smoothing technique to condition the trajectories on the heterogeneous data. This allows to integrate the available information in the computer model inputs, and to perform a diagnostics sanity check whereby confirming that data and simulation code are homogeneous. Once the prognostics window is reached, the same steps are applied, and then the
RUL distribution is estimated empirically. This allows to guide decision with regards to replacement or maintenance planning of the system. If prior to acting the decision there is a new data point acquired, the steps can be reapplied to eventually correct the RUL mean and the subsequent planning.

Using data assimilation mathematical terminology, let $\bm{Z} = (\bm{X},\bm{Y})$ be the state vector \citep{evensen-vossepoel-2022}, where $\bm{Y} = g(\bm{X})$ are the model outputs over a given time window, and $\bm{X}$ are the uncertain, time-invariant parameters. The objective is to estimate the joint posterior $p(\bm{Z} | \mathcal{D})$. We do so by using a modular approach:
\begin{itemize}
    \item Perform Bayesian model updating (BMU) to obtain the posterior distribution $h$ for the input parameters: $(\bm{X}|\mathcal{D}) \sim h_{\mathcal{D}}$.
    \item Apply smoothing to estimate $p(\bm{Y} | \mathcal{D}, \bm{X}\sim h_{\mathcal{D}})$, yielding the assimilated posterior approximation:
    \begin{equation}
        p(\bm{Z}|\mathcal{D}) \simeq h_{\mathcal{D}}(\bm{X})p(\bm{Y}|\mathcal{D},\bm{X}\sim h_{\mathcal{D}}).
    \end{equation}
\end{itemize}
 Once the prognostics window is reached, then the RUL distribution can be computed using the estimated density:
 \begin{equation}
    \mathbb{P}(\text{RUL}(D)\leq t_{j}|\mathcal{D}) = \int_{\mathbb{R}}\bm{1}\{\text{pr}_{j+1}(\bm{y})\geq D\}p(\bm{y}|\mathcal{D},\bm{X}\sim h_{\mathcal{D}})d\bm{y},
 \end{equation}
where the integral is estimated using a Monte Carlo ensemble $\{(\bm{X}^{(i)}, g(\bm{X}^{(i)}))\}_{i=1}^{n}\sim h_{\mathcal{D}}\otimes g\#h_{\mathcal{D}}$:
\begin{equation}
    \mathbb{P}(\text{RUL}(D)\leq t_{j}|\mathcal{D}) \approx \frac{1}{n}\sum_{i=1}^{n}\bm{1}\{\text{pr}_{j+1}\circ g(\bm{X}^{(i)})\geq D\}.
\end{equation}

\subsection{Bayesian model updating step}
\label{subsec23}
\begin{figure}[h!]
    \centering
    \includegraphics[width=1.0\textwidth]{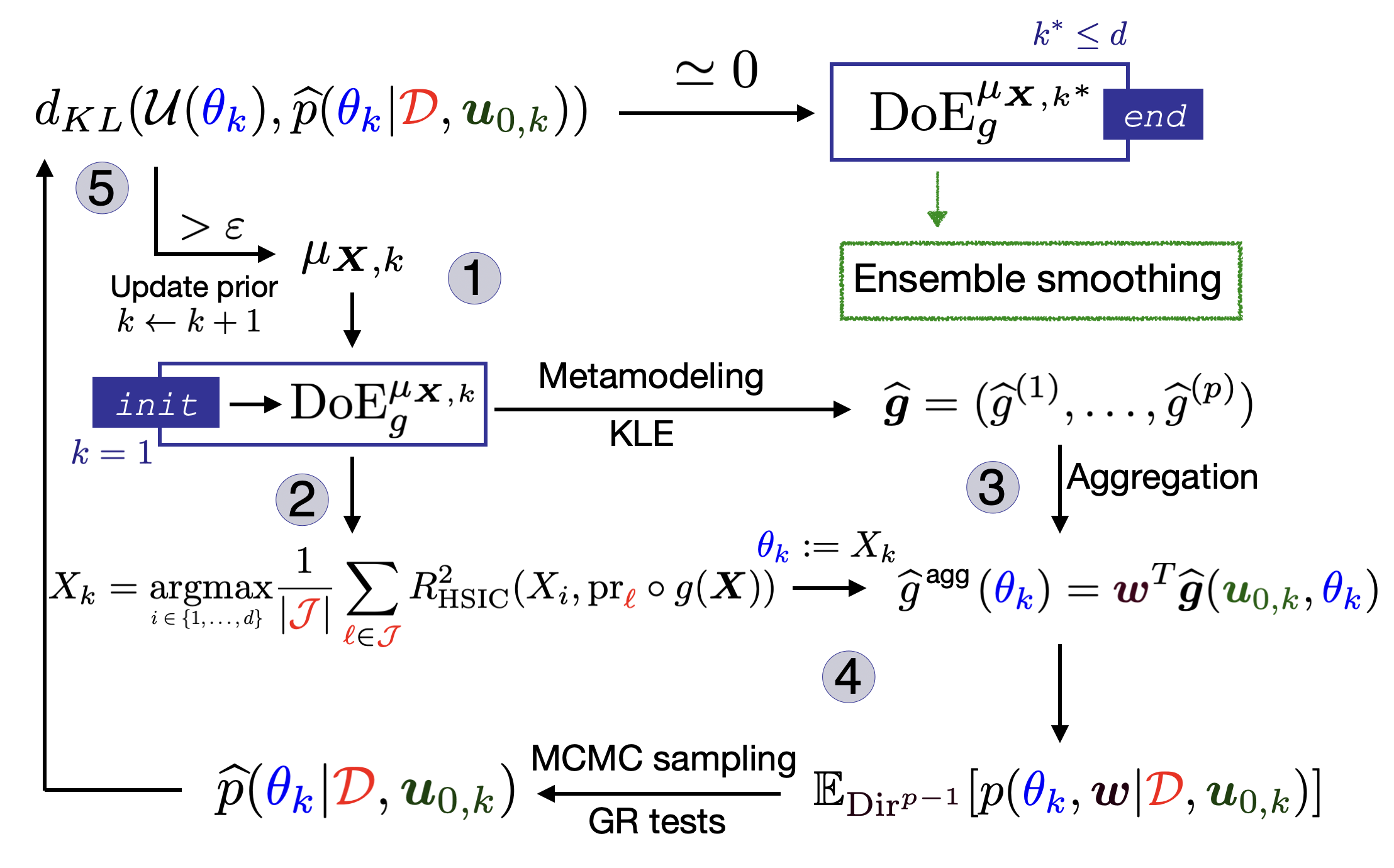}
    \caption{Proposed methodology for fusion of heterogeneous data in degradation prognostics.}
    \label{fig:general_scheme}
\end{figure}

We consider the input space $\mathcal{X} = \mathcal{X}_{1}\times\ldots\times \mathcal{X}_{d} \subset \mathbb{R}^d$ for the simulation model $g$. To minimize prior assumptions, we use a non-informative uniform prior with independent marginals: $\mu_{\bm{X}} = \mathcal{U}(\mathcal{X}_{1})\otimes\ldots\otimes\mathcal{U}(\mathcal{X}_{d})$. In most industrial applications it is rarely the case that one knows a prior dependency structure on the inputs, therefore in the proposed methodology independency is assumed from the beginning. Sampling from this prior yields a design of experiments $\text{DoE}^{\mu_{\bm{X}}}_{g} = \{(\bm{X}^{(i)}, g(\bm{X}^{(i)}))\}_{i=1}^{n}$, where $n$ is the sample size. Since the simulation model may be computationally expensive, we construct $p$ surrogate models $\widehat{\bm{g}} = (\widehat{g}^{\;(1)},\ldots,\widehat{g}^{\;(p)})$ using the DoE, trained for instance with different input hyperparameters or optimization procedures. These surrogates, built via supervised learning techniques such as polynomial chaos expansions (PCEs) \citep{sudret_2014, elmocayd_et_al_2018, jaber_et_al_2023}, Gaussian processes (GPs) \citep{rasmussen2006}, or feed-forward neural networks (ANNs) \citep{tripathy_et_al_2018, lefebvre_et_al_2023}, approximate $g$ as functions $\widehat{g}:\mathcal{X}\to\mathbb{R}^{N}$. Since no single surrogate is universally optimal, we aggregate the $p$ models into a single robust predictor $\widehat{g}^{\;\text{agg}}$ using a weighted average on the simplex $\Delta^{p-1}$, following expert aggregation strategies \citep{cesa-bianchi_prediction_2006}:
\begin{equation}
    \widehat{g}^{\text{agg}}(\bm{X}) = \sum_{i=1}^{p}w_{i}\widehat{g}^{\;(i)}(\bm{X}) = \bm{w}^{\top}\bm{\widehat{g}}(\bm{X}).
\end{equation}
The weights in $\bm{w}$ are treated as hyperparameters. Assuming independent marginals in $\bm{X}$, we iteratively assess each marginal's contribution to output uncertainty to update the most influential priors using heterogeneous data. The methodology proceeds for up to $d$ iterations (input dimension), as summarized in Figure~\ref{fig:general_scheme}:

\begin{enumerate}
    \item Initialize with $\mu_{\bm{X},0} = \mathcal{U}(\mathcal{X}_{1})\otimes\ldots\otimes\mathcal{U}(\mathcal{X}_{d})$ and generate the initial DoE.
    \item Rank variable influence using a kernel-based sensitivity analysis (Hilbert-Schmidt Independence Criterion, HSIC), averaged over all data time instances. Select the most influential variable $\theta_{k} := X_{k}$ at iteration $k$, fixing the remaining variables $\bm{U}_{k} = \bm{u}_{0,k}$ at nominal values.
    \item Build and aggregate surrogate models as described above.
    \item For $q$ heterogeneous data groups $\bm{y}^{1},\ldots,\bm{y}^{q}$, each with acquisition times $\mathcal{J}_{i}$, use a Bayesian approach to derive the posterior of $\theta_{k}$ given the data. The posterior integrates noise variance uncertainty and surrogate weights via Monte Carlo. Sampling from $p(\theta_{k}|\bm{y}^{1},\ldots,\bm{y}^{q}, \bm{u}_{0,k})$ is performed using Random-Walk Metropolis-Hastings MCMC.
    \item Compute the Kullback-Leibler divergence between prior and posterior to quantify information gain. If the data sufficiently informs $\theta_{k}$, update its prior in $\mu_{\bm{X},k}$, regenerate the DoE, and retrain surrogates.
\end{enumerate}
This process repeats until the information gain becomes negligible. A reproducible \href{https://github.com/EdgarJaber/bayes-calibration-for-prognostics.git}{Github repository} provides Python code and application results. The mathematical background for each individual step is detailed in the following paragraphs.

\section{Mathematical background used}\label{sec3}
\subsection{Kernel-based global sensitivity analysis for input variable selection}
\label{subsec31}
Among various global sensitivity analysis measures, the Hilbert-Schmidt Independence Criterion (HSIC) is recognized as a powerful dependency measure for sensitivity analysis \citep{gretton2005,DaVeiga2015,DaVeiga_Gamboa_Iooss_Prieur_Book_2021}. HSIC uncovers dependency structures between each input variable $X_{j} \in \bm{X}\in\mathbb{R}^d$ and the output of the physical code $Z_{\ell} = \text{pr}_{\ell}\circ g(\bm{X}) \in \mathcal{Z}_{\ell}\subset \mathbb{R}$ at a certain time instance $t_{\ell}\in\mathcal{J}$. The fundamental idea behind HSIC is to compare the joint probability distribution $ \mathbb{P}_{X_{j},Z_{\ell}} $ of the couple $(X_{j}, Z_{\ell})$ with the product of their marginal distributions $ \mathbb{P}_{X_{j}}\mathbb{P}_{Z_{\ell}} $. This comparison is performed using generalized covariance operators in reproducing kernel Hilbert spaces (RKHSs) \citep{Berlinet_Thomas_Agnan_BOOK}. A key property of HSIC is that a higher index value for a given variable indicates a stronger dependency with the output, this property in question is especially useful for variable selection \citep{DaVeiga2015}.
Let $i\in\{1,\ldots,d\}$ be an input parameter index and $k\in\mathcal{J}$ be a data time instance. Consider two RKHSs, $ \mathcal{F}_{i} $ and $ \mathcal{G}_{\ell} $, with respective characteristic kernels $ \kappa_{i} $ and $ \kappa_{\ell} $ \citep{Sriperumbudur_et_al_JMLR_2011}.
On the product space $ \mathcal{F}_{i} \times \mathcal{G}_{\ell} $, let's consider the product kernel $ \nu_{j\ell} = \kappa_{j}\otimes\kappa_{k} $, for all $ (X_{j},X_{j}') \in \mathcal{X}^{2}_{i}, (Z_{\ell},Z_{\ell}') \in \mathcal{Z}_{k}^{2} $ such as:
\begin{equation}
    \nu_{jk}\left((X_j,Z_\ell), (X_{j}',Z_{\ell}')\right) = \kappa_{i}(X_{j},X_{j}')\kappa_{k}(Z_{\ell},Z_{\ell}').
\end{equation} 
Then, the generalized covariance operator is defined as:
\begin{equation}
    C_{X_{j}Z_{\ell}} = \mathbb{E}_{(X_{j},Z_{\ell})}\left[\nu_{jk}\left((X_{j},Z_{k}), .)\right)\right] - \mathbb{E}_{X_{j}}\mathbb{E}_{Z_{k}}\left[\nu_{jk}\left((X_{j},Z_{\ell}), .)\right)\right],
    \label{eq:covariance_operator}
\end{equation}
and the HSIC measure quantifying dependence is given by the Hilbert-Schmidt norm of the covariance operator:
\begin{equation}
    \text{HSIC}(X_{j},Z_{\ell}) = \lVert C_{X_{j}Z_{\ell}} \rVert^{2}_{\text{HS}} = \text{Tr}(C_{X_{j}Z_{\ell}}^{\top}C_{X_{j}Z_{\ell}}).
\end{equation}
A fundamental result, proven in \citep{gretton2005,DaVeiga2015}, states that, for two characteristic kernels:
\begin{equation}
    \text{HSIC}(X_{j},Z_{\ell}) = 0 \Longleftrightarrow X_j \perp Z_{\ell}.
\end{equation}
As a counterpart, a nonzero HSIC value indicates a degree of dependence between the input $X_{j}$ and the output $Z_{\ell}$, justifying its use in global sensitivity analysis.
Estimating HSIC from data $\{(X^{(j)}_{i}, Z^{(j)}_{\ell})\}_{j=1}^{n}$ requires statistical approximations. Usually, two kinds of estimators exist, notably the U-statistics and the V-statistics \citep[Ch.~6]{DaVeiga_Gamboa_Iooss_Prieur_Book_2021}. A commonly used estimator formulates as:
\begin{equation}
    \widehat{\text{HSIC}}(X_{j}, Z_{\ell}) = \frac{1}{n^2}\text{Tr}\left(\mathbf{L}_{j}\mathbf{H}\mathbf{L}_{k}\mathbf{H}\right),
    \label{eq:hsic_estimator}
\end{equation}
where $ \mathbf{L}_{j}, \mathbf{L}_{k} $ are Gram matrices with entries $ (\mathbf{L}_{j})_{pq} = \kappa_{i}(X^{(p)}_{j}, X^{(q)}_{j}) $ and $ (\mathbf{L}_{k})_{pq} = \kappa_{k}(Y^{(p)}_{k}, Y^{(q)}_{k}) $ for $p,q\in\{1,\ldots,n\}$. The matrix $ \mathbf{H} $ is a centering matrix $ (\mathbf{H})_{pq} = \delta_{p,q} - \frac{1}{n} $ (here $\delta_{p,q}$ is the Dirac delta function). In practical usage for real-valued random variables, we use the Gaussian kernel parameterized by the empirical standard-deviation $\widehat{\sigma}_{X_{j}}$ of the input sample:
\begin{equation}
    \kappa_{j}(X^{(p)}_{j},X^{(q)}_{j}) = \exp\left(|X^{(p)}_{j} - X^{(q)}_{j}|^{2}/\widehat{\sigma}_{X_{j}}\right).
\end{equation}
In order to better interpret the ranking results, a normalized version of the index \citep{Marrel2021} is often considered:
\begin{equation}
    R^{2}_{\text{HSIC}}(X_{j},Z_{\ell}) = \frac{ \widehat{\text{HSIC}}(X_{j}, Z_{\ell})}{\sqrt{ \widehat{\text{HSIC}}(X_{j}, X_{j}) \widehat{\text{HSIC}}(Z_{\ell}, Z_{\ell})}} \in [0,1].
    \label{eq:r2_hsic}
\end{equation}
In the variable selection step of the iterative methodology proposed in Figure~\ref{fig:general_scheme}, the normalized $R^{2}_{\text{HSIC}}$ index is computed for all variables $X_{j}$ in $\bm{X}$. This index is computed only at the projections on all the data time instances $t_{\ell} \in \mathcal{J}$, without taking into account the heterogeneity of the data groups. If the time instances are not present in the predefined code time grid, we make use of interpolation operators to obtain an estimation of the output simulation. We average the corresponding $R^{2}_{\text{HSIC}}$ over all data time instances $\ell\in\mathcal{J}$ and then select the input $X_{k}$ with maximum sensitivity index:

\subsection{Surrogate modeling and validation strategies}
\label{subsec32}
The use of surrogate models is necessary to speed up the forward model simulations which is crucial in particular for MCMC methods, usually requiring a high number of calls. The latter come in different flavors, depending on whether an intrusive or a non-intrusive strategy is adopted \citep{leMaitre2012}. Intrusive methods also fall in the category of reduced order models (ROMs) with methods such as Galerkin expansions \citep{Quarteroni2016}. This method knows the underlying partial differential equation system to be solved as well as discretization and numerical method strategy, and with the use of snapshots (i.e., results of the computer model at different time-instances or unitary calls for different inputs), uses reduced bases to estimate the response variable of the physical model and thus the solution manifold \citep{benner2015}. In case where the numerical method used is not available, we can only work with the given-data snapshots of computer model $g:\mathcal{X}\to \mathbb{R}^N$, based on the input DoE $\{\bm{X}^{(i)}\}_{i=1}^{n} \sim \mu_{\bm{X}}$. In this case, the common strategy is to work with field surrogate models based on the Karhunèn-Loève expansions whose theory is briefly recalled hereafter.

\subsection{Karhunèn-Loève-based functions}
\label{53}
We consider a time-dependent simulation model $g: \mathcal{X} \rightarrow \mathbb{R}^N$, where each output $g(\bm{X}) = (g(t_1,\bm{X}), \ldots, g(t_N,\bm{X}))$ is a trajectory discretized on a regular time grid $\{t_1, \ldots, t_N\}$. The model is evaluated at $n$ input samples $\{\bm{X}^{(i)}\}_{i=1}^n \sim \mu_{\bm{X}}$, resulting in an output data matrix:
\begin{equation}
\bm{Y} = \left[ g(\bm{X}^{(1)}), \ldots, g(\bm{X}^{(n)}) \right] \in \mathbb{R}^{N \times n}.
\end{equation}
To reduce the dimensionality of the output space, we apply a Karhunèn–Loève expansion (KLE). First one needs to consider the empirical covariance matrix:
\begin{equation}
\hat{C} = \frac{1}{n} \bm{Y} \bm{Y}^\top.
\end{equation}
Then, a singular value decomposition (SVD) of $\bm{Y}$ is performed leading to:
\begin{equation}
\bm{Y} = \bm{V} \bm{\Sigma} \bm{W}^\top,
\end{equation}
where $\bm{V} = (\bm{\Phi}_{1},\ldots,\bm{\Phi}_{m})^{\top}\in \mathbb{R}^{N \times m}$ contains the Karhunèn-Loève orthonormal eigenvectors (also known as KLE modes), $\bm{\Sigma} = \mathrm{diag}(\sigma_1, \ldots, \sigma_m)$ with $\sigma_i^2$ the empirical output variances, $\sigma_{1}\geq \ldots \geq \sigma_{m}\geq 0$ and $m \leq \min(N,n)$ is the number of retained modes (or truncation level). The truncation level $m$ is chosen to capture a prescribed proportion of total variance, explained by the KLE. Each trajectory $g(\bm{X}^{(i)})$ is projected onto the retained KLE modes:
\begin{equation}
\xi_k(\bm{X}^{(i)}) = \langle g(\bm{X}^{(i)}), \bm{\Phi}_k \rangle = g(\bm{X}^{(i)})^\top \bm{\Phi}_k.
\label{eq:kl_modes}
\end{equation}
This yields $m$ scalar-valued datasets:
\begin{equation}
\text{DoE}_k = \left\{ \left( \bm{X}^{(i)}, \xi_k(\bm{X}^{(i)}) \right) \right\}_{i=1}^n, \quad k = 1, \ldots, m.
\end{equation}
For each mode $k$, we construct a surrogate model $\widehat{\xi}_k(\bm{X})$, typically, a Gaussian process or a polynomial chaos expansion. The full trajectory is then reconstructed as:
\begin{equation}
\widehat{g}(\bm{X}) = \sum_{k=1}^m \widehat{\xi}_k(\bm{X}) \bm{\Phi}_k.
\end{equation}
This KLE-based surrogate modeling procedure reduces the computational cost of the forward model and is particularly well-suited for Bayesian inference tasks involving such functional outputs. In order to validate the metamodel, we usually proceed by computing the predictivity coefficient $Q^2$ at the specific time-steps. Since in practice it is possible to use surrogates with different priors (i.e. various prior mean functions or covariance kernel on the GP \citep{rasmussen2006, Marrel2021} or prior degree and $q$-norm for the PCE \citep{jaber_et_al_2023}) but that give equally good predictive means, we choose to average the $p$ surrogates $(\widehat{g}^{(1)}, \ldots, \widehat{g}^{(p)})^{\top}$ by using a convex combination $(w_1, \ldots, w_p)^{\top} \in \Delta^{p-1}$, similar to an expert aggregation technique \citep{cesa-bianchi_prediction_2006}: 
\begin{equation}
\widehat{g}^{\text{agg}}(\bm{X}) = \sum_{i=1}^{p}w_{i}\widehat{g}^{(i)}(\bm{X}).
\end{equation}
This approach is useful for a couple of reasons. Firstly, since individual surrogates may have biases due to their specific assumptions or hyperparameters, it allows to reduce the total bias by averaging it out. Secondly, it increases the stability of the model since the combined prediction is less sensitive to the misspecification of any single surrogate, which leads to more stable and generalizable results. Finally since there is no best surrogate in all generality, aggregation is a suitable way to incorporate all the available plausible models. Model bias is treated in the classical Bayesian calibration approach proposed by \citep{kennedy_ohagan2001} by adding an additional stochastic process in the second equation in equation ~\eqref{eq:bayes_update} and calibrating the hyperparameters involved. However this is prone to misspecification since there are no guarantees for an informed choice for its prior covariance. Our method is more accurate due to the consensus of several well-performing models rather than relying on a single potentially suboptimal choice.

\subsection{Heterogeneous data groups}
\label{subsec33}
In some engineering prognostics applications, relevant degradation data can be collected through multiple, distinct methods. For example, direct measurements may be obtained from sensors that monitor the degradation indicator, while indirect methods may use correlated quantities and regression models—developed through feature engineering—to estimate the degradation state. Additionally, the available data often exhibit both temporal and structural heterogeneity: measurements may be acquired at irregular, non-periodic time points that do not necessarily align with the simulation model’s time grid, and the data may be sparse in time or space. To capture this, we consider $q$ heterogeneous data groups, denoted as $\bm{y}^{1}, \ldots, \bm{y}^{q}$, where each group is associated with its own acquisition time grid $\mathcal{J}_{i} = \{ t^{i}_{1},\ldots,t^{i}_{m_{i}}\}$. The union of all acquisition times is $\mathcal{J} = \cup_{i=1}^{q}\mathcal{J}_{i}$, with total cardinality $|\mathcal{J}| = m_{1}+\ldots + m_{q}$. For each group $i$ and time index $j=1,\ldots,m_{i}$, the observed data are modeled as:
\begin{equation}
    y^{i}(t^{i}_{j}) = \delta(t^{i}_{j}) + \eta^{i}_{j},
\end{equation}
where $\delta$ is the true (latent) degradation process and $\eta^{i}_{j} \sim \mathcal{N}(0,\sigma^{2}_{i})$ represents additive Gaussian noise. We assume the noise is homoskedastic within each group (constant variance $\sigma^2_i$ for group $i$), but heteroskedastic across groups (different variances between groups). This modeling approach allows us to flexibly incorporate multiple, heterogeneous sources of information, each with its own noise characteristics and sampling schedule, into the data fusion framework.

\subsection{Bayesian data fusion}
\label{subsec34}
As explained in the introduction, the state-space model variability is given all at once through a pushforward of the uncertain input variables independent probabilistic model. The problem is to estimate the posterior distribution after considering the heterogeneous data of the most influential input variable at iteration $k\leq d$. Assume that there is some value $\theta_{k}$ such that $\delta(t^{i}_{j}) = g(t^{i}_{j}, \bm{u}_{0,k}, \theta_{k})$, where $\theta_{k}$ is defined in eq. \eqref{eq:max_hsic} and $\bm{u}_{0,k}$ is the nominal value of the $d-1$ variables at step $k$. We therefore have:
\begin{equation}
    y^{i}(t^{i}_{j}) = g(t^{i}_{j},\bm{u}_{0,k}, \theta_{k}) + \eta^{i}_{j},\;\;\;\forall j\in\mathcal{J}_{i}.
\end{equation}
Usually $\mathcal{J}_{i}$ is a coarser grid than the original time grid of the simulation model. Therefore, we need to interpolate the computer code output. As an interpolator operator, one can choose $\bm{\mathcal{G}}_{i}: \mathcal{X} \to\mathbb{R}^{|\mathcal{J}_{i}|}$ to interpolate the simulation model onto the time-grid of the data points. The choice of this interpolator is driven by prior regularity properties of the degradation trajectory such as monotonicity, which is known since the computer code is treated as a grey-box. Note that this hypothesis is usually verified in degradation applications including the ones we present. We drop the dependence on the latent variable in the rest of this paragraph, it is important to note however that the results are conditioned on these latent variables, as shown in \ref{fig:general_scheme}. In vector form, the problem is recast as:
\begin{equation}
    \bm{y}^{i} = \bm{\mathcal{G}}_{i}(\theta_{k}) + \bm{\eta}^{i}, \;\forall i\in\{1,\ldots,q\},
\end{equation}
with $\bm{\eta}^{i} \sim \mathcal{N}(0, \sigma^{2}_{\bm{\eta}^{i}})$, where the covariance matrix of the noise is defined as $\sigma^{2}_{\bm{\eta}^{i}} :=\sigma_{i}^{2}I_{m_{i}}$. To simplify the presentation, it is assumed now that only one data group $\bm{y} \in \mathbb{R}^{m}$ is considered. The goal is to use the Bayes' theorem to estimate the posterior distribution of the influential input variable $\theta_{k}$ given the data:
\begin{equation}
    p(\theta_{k}|\bm{y}) \propto p(\theta_{k})p(\bm{y}|\theta_{k}).
\end{equation}
This likelihood quantifies how well parameter $\theta_{k}$ explains the observed data. Assuming Gaussian residuals that are independent of $\theta_{k}$, the likelihood is Gaussian with noise variance $\sigma^2_{\bm{\eta}}$ (homoskedastic case). The posterior density is then:
\begin{equation}
    p(\theta_{k},\sigma^{2}_{\bm{\eta}}|\bm{y}) \propto p(\theta_{k},\sigma^{2}_{\bm{\eta}})\, (\sigma^{2}_{\bm{\eta}})^{-m/2} \exp\left(-\frac{1}{2\sigma^{2}_{\bm{\eta}}}\lVert \bm{y} - \bm{\mathcal{G}}(\theta_{k})\rVert^{2}\right),
    \label{eq:posterior_proba}
\end{equation}
where $m$ is the number of data points. At this stage, it is possible to numerically integrate the standard deviation of the noise and modify the posterior distribution, but this often leads to numerical instabilities in MCMC chains. A carefully chosen prior on $\sigma^{2}_{\bm{\eta}}$ allows for analytical error integration using Bayes' theorem. In the present paper, we propose an extension of the proposition found in \citep{keller2022} to multiple heterogeneous (heteroskedastic) groups of data:

\begin{Proposition}{}
Assume $\lambda := 1/ \sigma_{\bm{\eta}}^{2}\sim \mathcal{G}( \frac{m}{2},\frac{1}{2}\lVert \bm{y} - \bm{\mathcal{G}}(\theta_{k}) \rVert^{2})$, where $m$ is the number of data points in $\bm{y}$; $\theta \sim \mathcal{U}(\mathcal{X}_{*})$, and $p(\theta,\lambda) \propto \lambda^{-1}$. Then:
\begin{equation}
    p(\theta_{k}|\bm{y})\propto\lVert \bm{y} - \bm{\mathcal{G}}(\theta_{k})\rVert^{-m}.
\end{equation}
Moreover, if multiple groups of data at different time-instances are considered for assimilation, then, one has $\bm{y}^{1},\ldots,\bm{y}^{q}$, with respective priors on the inverse of their standard deviations $\lambda_{i}\sim \mathcal{G}(\frac{m_{i}}{2}, \frac{1}{2}\lVert\bm{y}^{i} - \bm{\mathcal{G}}_{i}(\theta_{k})\rVert^{2})$, and the generalization thus reduces to:
\begin{equation}
    p(\theta_{k}|\bm{y}^{1},\ldots,\bm{y}^{m})\propto \prod_{i=1}^{q}\lVert \bm{y}^{i} - \bm{\mathcal{G}}_{i}(\theta_{k})\rVert^{-m_{i}}.
    \label{eq:posterior_multifidelity}
\end{equation}
\label{prop1}
\end{Proposition}
\vspace{-3mm}
\noindent The proof is provided in Appendix \ref{subsec72}. To be exact, the posterior distribution obtained is also conditioned on the fixed latent variables, thus we have $p(\theta_{k}|\bm{y}^1,\ldots,\bm{y}^q,\bm{u}_{0,k})$, and since we use an aggregation of $p$ valid surrogates of the original degradation simulation model, we induce additional hyperparameters that we consider following a $\text{Dir}^{p-1}$ distribution. Thus in eq. ~\eqref{eq:posterior_multifidelity}, one can rewrite the posterior as:
\begin{equation}
    p(\theta_{k}, \bm{w}|\bm{y}^1,\ldots,\bm{y}^q,\bm{u}_{0,k}) \propto \prod_{i=1}^{q}\lVert \bm{y}^{i} - \langle \bm{w}, \bm{\mathcal{G}}^{\text{agg}}_{i}(\theta_{k})\rangle \rVert^{-m_{i}},
\end{equation}
where $\bm{\mathcal{G}}^{\text{agg}}_{i}(\theta_{k}) := (\bm{\mathcal{G}}_{i,1}(\theta_{k}),\ldots,\bm{\mathcal{G}}_{i,p}(\theta_{k}))$. At this stage we marginalize on the weights to obtain the posterior distribution of the influential input variable $\theta_{k}$ given the heterogeneous data as well as the aggregate surrogates:
\begin{equation}
    p(\theta_{k}|\bm{y}^1,\ldots,\bm{y}^q,\bm{u}_{0,k}) \propto \int_{\Delta^{p-1}}\prod_{i=1}^{q}\lVert \bm{y}^{i} - \langle \bm{w}, \bm{\mathcal{G}}^{\text{agg}}_{i}(\theta_{k})\rangle \rVert^{-m_{i}}d\bm{w}.
\end{equation}
Then, a Monte Carlo approximation of the integral is performed, amounting to sampling $\bm{w}^{(1)},\ldots,\bm{w}^{(M)}$ from the Dirichlet distribution $\text{Dir}^{p-1}$ and computing the average of the likelihood functions:
\begin{equation}
    p(\theta_{k}|\bm{y}^1,\ldots,\bm{y}^q,\bm{u}_{0,k}) \propto \frac{1}{N}\sum_{i=1}^{M}\prod_{j=1}^{q}\lVert \bm{y}^{j} - \langle \bm{w}^{(i)}, \bm{\mathcal{G}}^{\text{agg}}_{j}(\theta_{k})\rangle \rVert^{-m_{j}}.
\end{equation}
Since for numerical purposes we will consider the log-likelihood function, $\log p(\theta_{k}|\bm{y}^1,\ldots,\bm{y}^q,\bm{u}_{0,k})$, then we will use the log-sum-exp trick in order to avoid numerical instabilities. The log-likelihood is rewritten as:
\begin{equation}
    \begin{aligned}
    \log p(\theta_{k}|\bm{y}^1,\ldots,\bm{y}^q,\bm{u}_{0,k}) &\propto \log \sum_{i=1}^{M} \exp\left(-\sum_{j=1}^{q}m_{j}\log(\lVert \bm{y}^{j} - \langle \bm{w}^{(i)}, \bm{\mathcal{G}}^{\text{agg}}_{i}(\theta_{k})\rangle \rVert )- C\right) + C,
    \end{aligned}
\end{equation}
where $C := \max_{i=1,\ldots,M} \sum_{j=1}^{q}m_{j}\left(\lVert \bm{y}^{j} - \langle \bm{w}^{(i)}, \bm{\mathcal{G}}^{\text{agg}}_{i}(\theta_{k})\rangle \rVert\right)$. The log-likelihood is then used in the MCMC algorithm to sample from the posterior distribution of the influential input variable $\theta_{k}$. This method cannot be used for integrating the latent variables uncertainties, since we have tried numerical experiments and have observed that this often averages the likelihood too much and the posterior distribution is not well estimated, especially when the dimension is important. The integration of the latent variables $\bm{u}_{0,k}$ remain an open-question with the current methodology and is left for future work.

\subsection{Random-Walk Metropolis-Hastings}
\label{subsec35}
The Random-Walk Metropolis-Hastings (RWMH) algorithm \citep{rubinstein2011} is a MCMC method used to sample from the log-posterior distribution $\log p(\theta_{k}|\bm{y}^1,\ldots,\bm{y}^q,\bm{u}_{0,k})$. Starting from an initial guess $\theta_{k}^{(0)}$, the algorithm iteratively generates a candidate $\theta_{k}^{*}$ from a proposal distribution $q(\theta_{k}^{*}|\theta_{k}^{(i)})$. In out application, we choose uniform proposals centered on the barycenter value of $\mathcal{X}_{*}$ with different step-sizes. The candidate is then accepted with probability:
\begin{align}
    \alpha = \min \Bigg\{ 1, \exp\Big[ &\log p(\theta_{k}^{*}|\bm{y}^1,\ldots,\bm{y}^q,\bm{u}_{0,k}) - \log p(\theta_{k}^{(i)}|\bm{y}^1,\ldots,\bm{y}^q,\bm{u}_{0,k}) \nonumber \\
    &\quad + \log q(\theta_{k}^{(i)}|\theta_{k}^{*}) - \log q(\theta_{k}^{*}|\theta_{k}^{(i)}) \Big] \Bigg\}.
\end{align}
If accepted, the chain moves such that $\theta_{k}^{(i+1)} = \theta_{k}^{*}$; otherwise we stay at the same place such that $\theta_{k}^{(i+1)} = \theta_{k}^{(i)}$. This process is repeated for a large number of iterations to ensure convergence to the target posterior distribution.  In order to test the convergence of the algorithm, the Gelman-Rubin diagnostic \citep{gelman1992} is used. To do so, $J$ Markov chains are initialized with different initial values and the RWMH algorithm is runned for a large number of iterations. After the burn-in phase, a chain $\theta_{k,1}^{(i)},\ldots,\theta_{k,L}^{(i)}$ for $i=1,\ldots,J$ is obtained where $L$ is the iteration number of the chain. The mean value of the chain and between chains are computed: 
\begin{equation}
    \overline{\theta}_{k,i} = \frac{1}{L}\sum_{j=1}^{L}\theta_{k,j}^{(i)},\;\;\overline{\theta}_{k,*}= \frac{1}{J}\sum_{i=1}^{J}\overline{\theta}_{k,i},
\end{equation}
as well as the variances of the means of the chains and the mean of the variances of one chain:
\begin{equation}
    B = \frac{L}{J-1}\sum_{i=1}^{J}(\overline{\theta}_{k,*} - \overline{\theta}_{k,i}^{2}),\;\; W = \frac{1}{J}\sum_{i=1}^{J}\left(\frac{1}{L-1}\sum_{j=1}^{L}(\theta_{k,j}^{(i)} -     \overline{\theta}_{k,j})^{2}\right).
\end{equation}
The Gelman-Rubin diagnostic is defined as:
\begin{equation}
    R = \frac{(1-1/L)W + (1/L)B}{W} \to 1,\; L\to\infty.
\end{equation}
The algorithm is considered to have converged when $R$ is close to 1 , i.e. for $L\to +\infty$. The resulting sample is exponentiated and renormalized to obtain a draw from an estimate of the desired posterior, and  also use kernel density estimation to obtain the functions.

\subsection{Information gain and Kullback-Leibler Divergence}
\label{subsec36}
The information gain (IG) quantifies the reduction in uncertainty about a random variable after observing data \citep{lindley1956}. In the context of Bayesian inference, it measures how much the posterior distribution $p(\theta_{k}|\bm{y}^1,\ldots,\bm{y}^q,\bm{u}_{0,k})$ differs from the prior uniform distribution $\mathcal{U}(\theta_{k})$. This difference is formally captured by the Kullback-Leibler (KL) divergence, defined as:
\begin{equation}
    d_{\text{KL}}\left(\mathcal{U}(\theta_{k}, p(\theta_{k}|\bm{y}^1,\ldots,\bm{y}^q,\bm{u}_{0,k}))\right) = \mathbb{E}_{(\theta_{k}|\bm{y}^1,\ldots,\bm{y}^q,\bm{u}_{0,k})}\log \left[\frac{p(\theta_{k}|\bm{y}^1,\ldots,\bm{y}^q,\bm{u}_{0,k})}{ \mathcal{U}(\theta_{k})}\right].
\end{equation}
A higher KL divergence indicates that the data significantly updates the prior, leading to a more concentrated posterior distribution. This metric is crucial in the iterative updating of the methodology presented in figure \ref{fig:general_scheme} as it helps to determining whether the assimilation of new data justifies modifying the prior distribution of the input variables. In numerical implementations, a good tradeoff value for the threshold of the Kullback-Leibler divergence was found to be $\epsilon = 0.1$. There are two possible outcomes of the algorithm: either the sensitivity assessment picks a variable whose prior has already been informed at a previous iteration, and therefore the Kullback-Leibler divergence is null, or it picks a variable that remains uninformed by the heterogeneous data, therefore posterior and prior are identical. 

\subsubsection{Ensemble Kalman Smoother (EnKS).}
The Ensemble Kalman Smoother (EnKS) \citep{evensen2000ensemble} is an extension of the Ensemble Kalman filter that assimilates all available observations to produce smoothed estimates of a system's state across all time steps. In the scalar, offline setting, we consider an ensemble of $n$ trajectories $\{g(\bm{X}^{(i)})\}_{i=1}^n$, generated from the simulation model (or from some surrogate of the simulation model). Let the $i$-th group of observations $\{y^{i}(t^{i}_{k})\}_{k=1}^{m_{i}}$ be available at a subset of the simulation times $\{t_{1},\ldots,t_{N}\}$, each corrupted by Gaussian noise with known variance $\sigma^{2}_{i}$.

Denoting for each time $t\in\{t_{1},\ldots,t_{N}\}$, the output of the simulation code as $g(t,\bm{X}) =: Y_{t}$, we define the ensemble mean $\overline{Y}_t = \frac{1}{n} \sum_{p=1}^N Y^{(p)}_t$ and ensemble anomalies $A_t^{(p)} = Y_t^{(p)} - \overline{Y}_t$. The cross-covariance matrix between the ensemble states at all times $t_{j}$ and the observed state at time $t^{i}_{k}$ is computed as
\begin{equation}
C_{kj} = \frac{1}{n - 1} \sum_{p=1}^n A_{t^{i}_k}^{(p)} A_{t_{j}}^{(p)}.
\end{equation}
The observation variance at $t^{i}_{k}$ is similarly estimated by $C_{kk} = \frac{1}{n - 1} \sum_{p=1}^N (A_{t^{i}_k}^{(p)})^2$. The scalar Kalman gain vector $K_t$ is then given by:
\begin{equation}
K_{t_{j}} = \frac{C_{kj}}{C_{kk} + \sigma^{2}_{i}}.
\end{equation}
The innovation term is defined as $d^{(p)} = y^{i}(t^{i}_k) - Y^{(p)}_{t^{i}_k}$, and the smoother update is applied to all ensemble members across all times $t\in\{t_{1},\ldots,t_{N}\}$:
\begin{equation}
Y_t^{(p)} \leftarrow Y_t^{(p)} + K_{t_{j}} \cdot d^{(p)}.
\end{equation}
This update is repeated for each observation time $t^{i}_k$, resulting in an ensemble of smoothed trajectories that incorporate the full observation sequence. The method assumes approximate linear-Gaussian behavior in the ensemble statistics and requires no model re-evaluation, making it suitable for post hoc assimilation.

\section{Numerical implementations}
\label{sec4}
\subsection{Numerical tools and reproductibility}
\label{subsec41}
For numerical applications, this paper presents two models of degradation prognostics:
\begin{itemize}
    \item \emph{Paris-Erdogan's law for crack growth}: An empirical model widely used in fracture mechanics to predict the rate of crack propagation in materials under cyclic loading. The model relates the crack growth rate to the range of the stress intensity factor and is parameterized by material constants and loading conditions. It serves as a benchmark for prognostics and we use it to test our BMU step of the methodology.
    \item \emph{Clogging simulation model for steam generators (THYC-Puffer-DEPO code)}: An industrial multiphysics simulation code developed by EDF R\&D for predicting the clogging kinetics in steam generators of pressurized water nuclear reactors. The model incorporates thermohydraulic, chemical, and particulate transport phenomena, and is used to forecast the evolution of clogging over the operational lifetime of the asset. We use this test-case to showcase the applicability of the full methodology, including the smoothing step.
\end{itemize}
The methodology and numerical experiments described in this paper are fully reproducible. All code and application results are available in the following GitHub repository: 
\begin{center}
\small{\url{https://github.com/EdgarJaber/bayes-calibration-for-prognostics.git}}
\end{center}
We use an optimized version of the RWMH code implemented in the OpenTURNS Python library \citep{baudin2017}, based on the Adaptive Metropolis algorithm to automatically tune the parameters of the Markov chain. It is also implemented in C++, allowing to have similar execution times to state of the art MCMC software such as STAN or JAGS.

\subsection{Fracture propagation model}
\label{subsec42}
Fracture propagation models are essential in predicting the growth of cracks in materials under stress. One of the widely used models for this purpose is Paris-Erdogan's empirical law \citep{paris1960} which describes the rate of crack growth per cycle of loading in terms of the stress intensity factor range. The model is particularly useful in the field of fatigue analysis, where it helps in estimating the RUL of components subjected to cyclic loading. By integrating the crack growth rate equation over the number of loading cycles, the total number of cycles to failure can be estimated. This information is crucial for maintenance planning and ensuring the safety and reliability of structures subjected to cyclic loading. This model is widely used as a prognostics benchmark in the field of fracture mechanics to predict the fatigue life of components as well as for Bayesian model updating \citep{zarate2012, peng2015}. 

\subsubsection{Paris-Erdogan's Law}
\label{subsubsection421}
Paris-Erdogan's law is an phenomenological relationship that relates the crack growth rate to the range of the stress intensity factor, $\Delta K$. The law is expressed as:
\begin{equation}
    \frac{da}{dN} = C (\Delta K)^m,
\end{equation}
where $a$ is the crack length, $N$ is the number of loading cycles, $C$ and $m$ are material constants that need to be determined experimentally, $\Delta K$ is the stress intensity factor range, defined as $\Delta K = K_{\text{max}} - K_{\text{min}}$, where $K_{\text{max}}$ and $K_{\text{min}}$ are the maximum and minimum stress intensity factors during a loading cycle. Exponent $m$ characterizes the material's resistance to crack growth, and coefficient $C$ represents the crack growth rate under a given stress state \citep{Bourinet2017}.

\subsubsection{Stress intensity factor}
\label{subsubsection422}

The stress intensity factor, $K$, is a measure of the stress state near the tip of a crack and is influenced by the applied load, crack size, and geometry of the component. For a mode I (opening mode) crack, the stress intensity factor is given by:
\begin{equation}
    K = \sigma \sqrt{\pi a} Y,
\end{equation}
where $\sigma$ is the applied stress, $a$ is the crack length and $Y$ is a dimensionless geometry factor that depends on the shape and size of the component and the crack. Therefore, the stress intensity factor range, $\Delta K$, can be expressed as:
\begin{equation}
    \Delta K = (\sigma_{M} - \sigma_{m})\sqrt{\pi a} Y.
\end{equation}

\subsubsection{Input probabilistic modeling}
\label{subsubsection423}
The uncertain input variables in the Paris-Erdogan's law fracture propagation model include the material constants $C$ and $m$, which characterize the material's resistance to crack growth under cyclic loading. The initial crack length $a(0)$ is tracked to predict the component's remaining life. The applied stresses $\sigma_{M}$ and $\sigma_{m}$ influence the stress intensity factor and the crack growth rate. Lastly, the geometry factor $Y$ accounts for the component's shape and its effect on the stress intensity factor. Uniform distributions with independent marginals are chosen here for the sake of illustration of the methodology, even though other distributions are chosen in practice for this model and work has established correlation between $C$ and $m$ (see \citep{Bourinet2017}).

\begin{table}[h!]
\centering
\begin{tabular}{|c|c|c|}
    \hline
    \textbf{Input Variable} & \textbf{Nominal value} & \textbf{Distribution} \\
    \hline
    $C$ & $C_0 = 1\times 10^{-10}$ & $\mathcal{U}[0.9\times C_{0}, 1.1\times C_{0}]$ \\
    $m$ & $m_{0} = 3$ & $\mathcal{U}[m_{0} - 0.1 m_{0} + 0.1]$ \\
    $\sigma_{M}$ &  $\sigma_{M, 0} = 100 \;(\text{MPa})$ & $\mathcal{U}[\sigma_{M, 0} - 5.0, \sigma_{M, 0} + 5.0]$ \\
    $\sigma_{m}$ & $\sigma_{m, 0} = 10 \;(\text{MPa})$ & $\mathcal{U}[\sigma_{m, 0} - 1.0, \sigma_{m, 0} + 1.0]$ \\
    $Y$ & $Y_0 = 1.1$ & $\mathcal{U}[Y_{0} - 0.01, Y_{0} + 0.01]$ \\
    $a(0)$ & $a_0 = 1\times 10^{-3} \;(\text{m})$ & $\mathcal{U}[a_0 - 1\times 10^{-4} , a_0 + 1\times 10^{-4}]$ \\
    \hline
\end{tabular}
\caption{Distributions of the Paris-Erdogan's law input variables.}
\end{table}

\subsubsection{Numerical results}
\label{subsubsection424}
\begin{figure}[h!]
    \centering
    \includegraphics[width=1.0\textwidth]{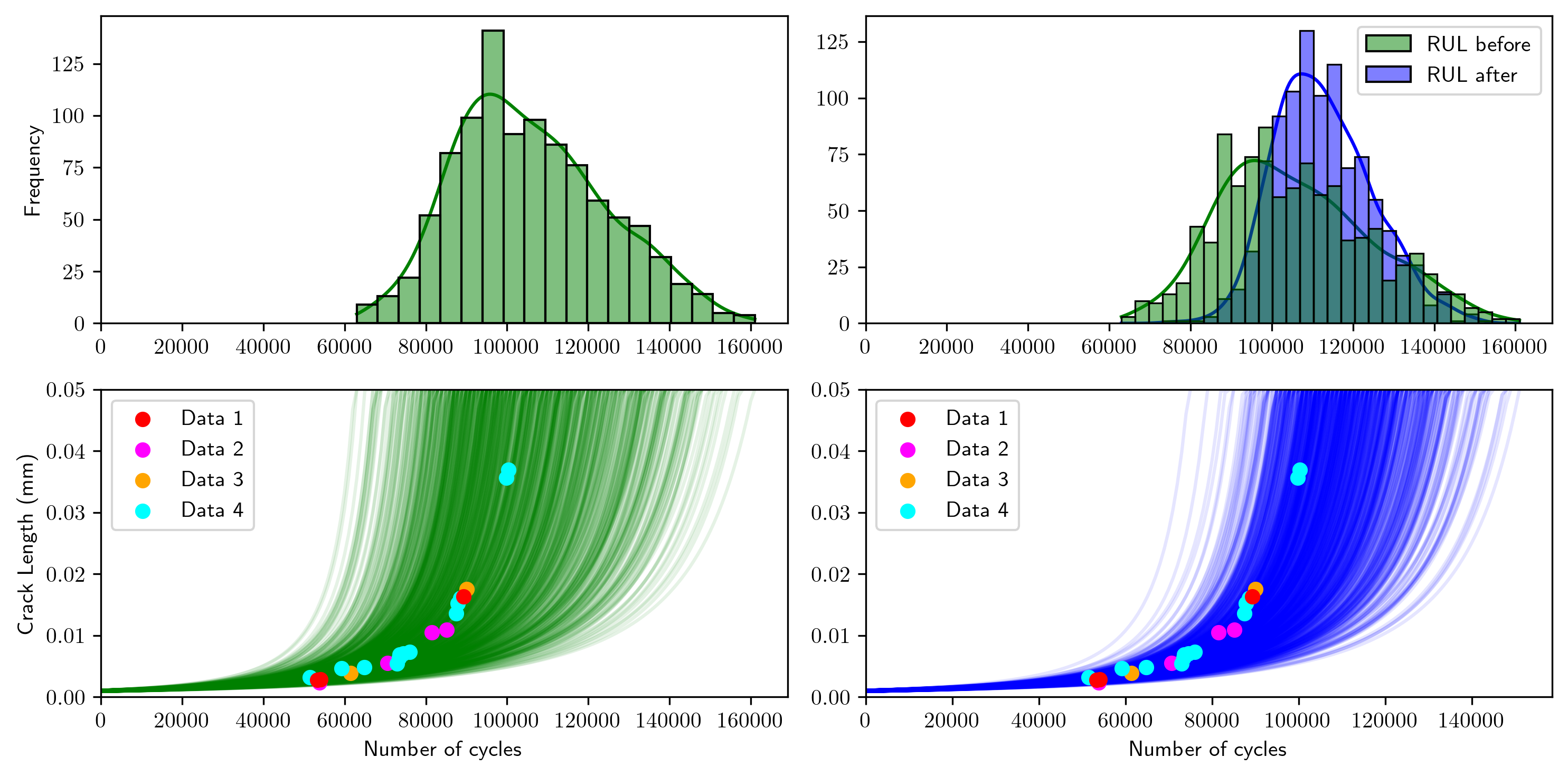}
    \caption{Monte Carlo sample of trajectories from the Paris-Erdogan's law before and after heterogeneous data fusion.}
    \label{fig:paris_law_uq}
\end{figure}

\begin{figure}[h!]
    \centering
    \includegraphics[width=1.0\textwidth]{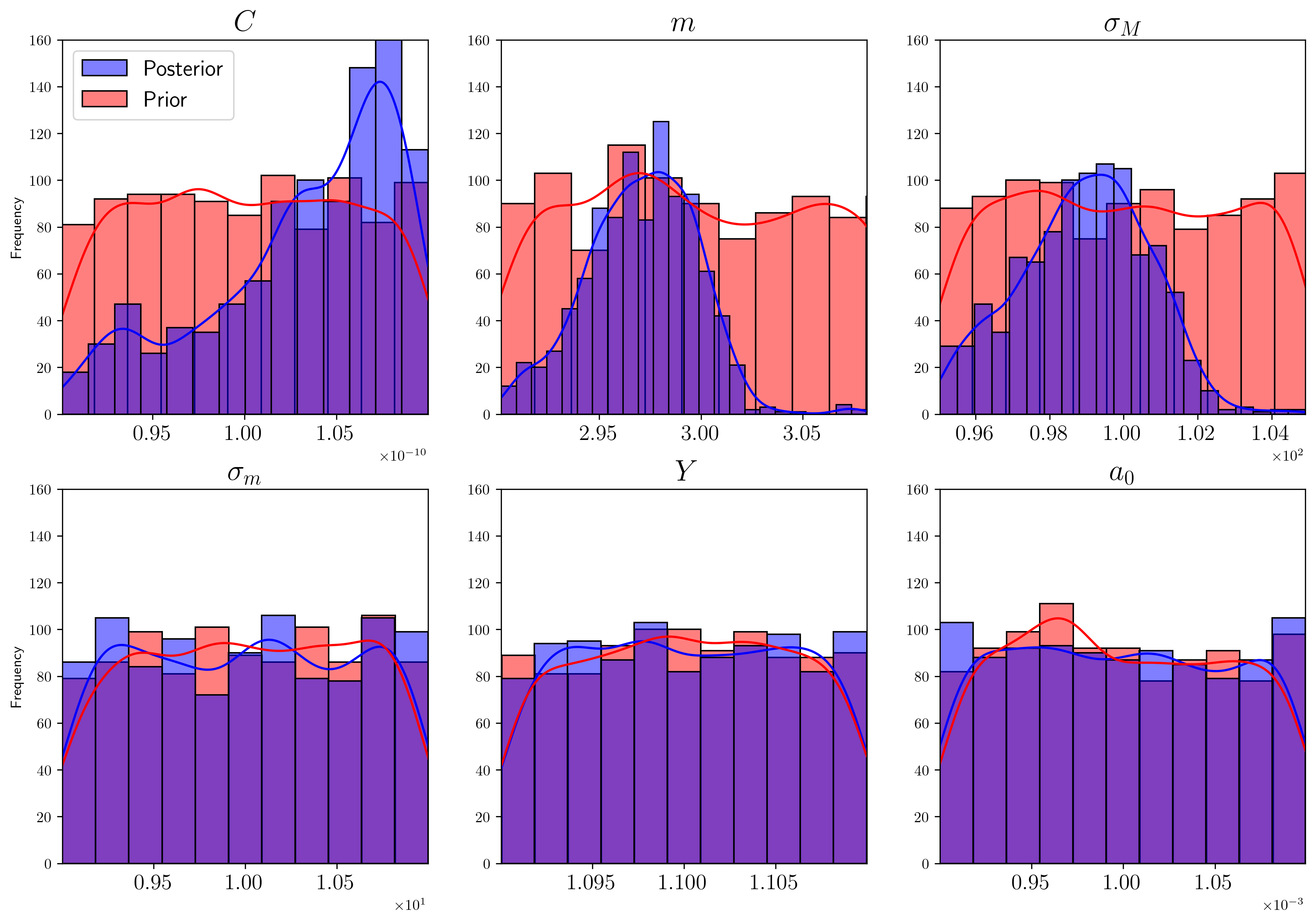}
    \caption{Prior and posterior distributions of the Paris-Erdogan's law input variables.}
    \label{fig:paris_posterior_distributions}
\end{figure}

In order to solve the crack size ordinary differential equation, an explicit Euler scheme of the Paris-Erdogan's law model is implemented, and set a RUL threshold at $D = 5$ cm. For illustration purposes, a series of $q=4$ different data groups are generated by perturbing a nominal degradation curve with different values of Gaussian noise. The data is also generated at different time instances. As shown in Figure~\ref{fig:paris_law_uq}, the $n=10^3$ Monte Carlo sample of trajectories from the Paris-Erdogan's law model show a wide range of possible outcomes of the RUL due to the uncertainty in the input variables. We start by interpolating linearly the resulting trajectories on the time instances of the data. Afterwards, we run the data fusion methodology presented to update the prior distributions of the input variables. Since the unitary call to the Euler scheme is not costly, there is no need for a metamodel here and we thus omit the integration on additional hyperparameters on $\Delta^{p-1}$. The prior and posterior distributions of the Paris-Erdogan's law input variables are shown in Figure~\ref{fig:paris_posterior_distributions}. The posterior distributions of parameters $C, m, \sigma_{M}$ are modified and more centered around a mean value compared to the prior uniform distributions that are homogeneous on the whole support, indicating that the data assimilation process has reduced the uncertainty in the input variables. However, the data does not inform variables $\sigma_{m}, Y$ and $a(0)$. We can see that the process has also reduced the uncertainty on the RUL, as the posterior distribution is better concentrated around a mean value. Therefore, updated distributions can be used to make more accurate predictions of the RUL of the component based on the observed data.

\subsection{Clogging of steam generators in pressurized water reactors}
\label{subsec43}
\begin{figure}[h!]
    \begin{minipage}{0.5\textwidth}
    \centering
        \includegraphics[width=1.0\textwidth]{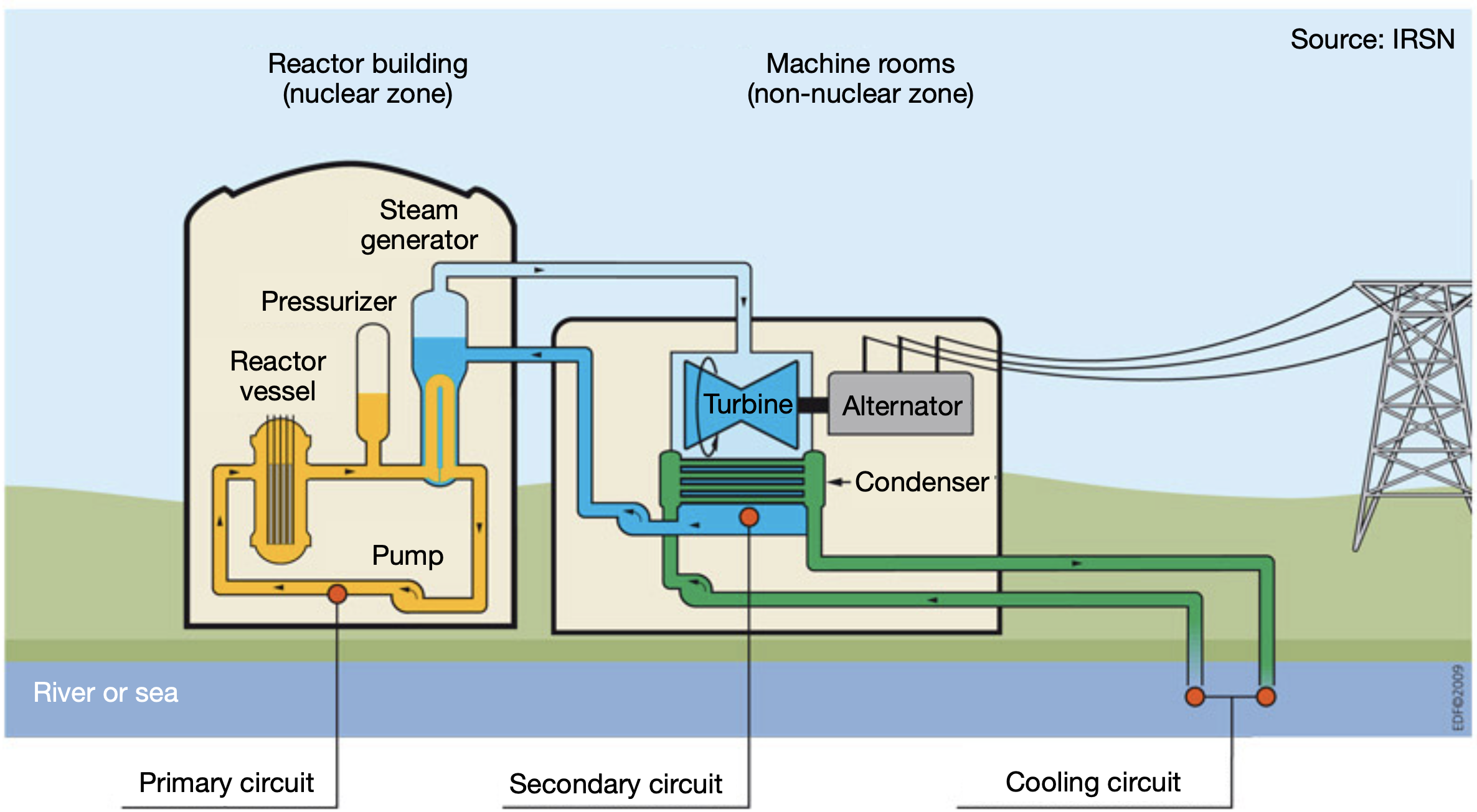} 
    \end{minipage}
    \begin{minipage}{0.25\textwidth}
    \centering
        \includegraphics[width=1.9\textwidth]{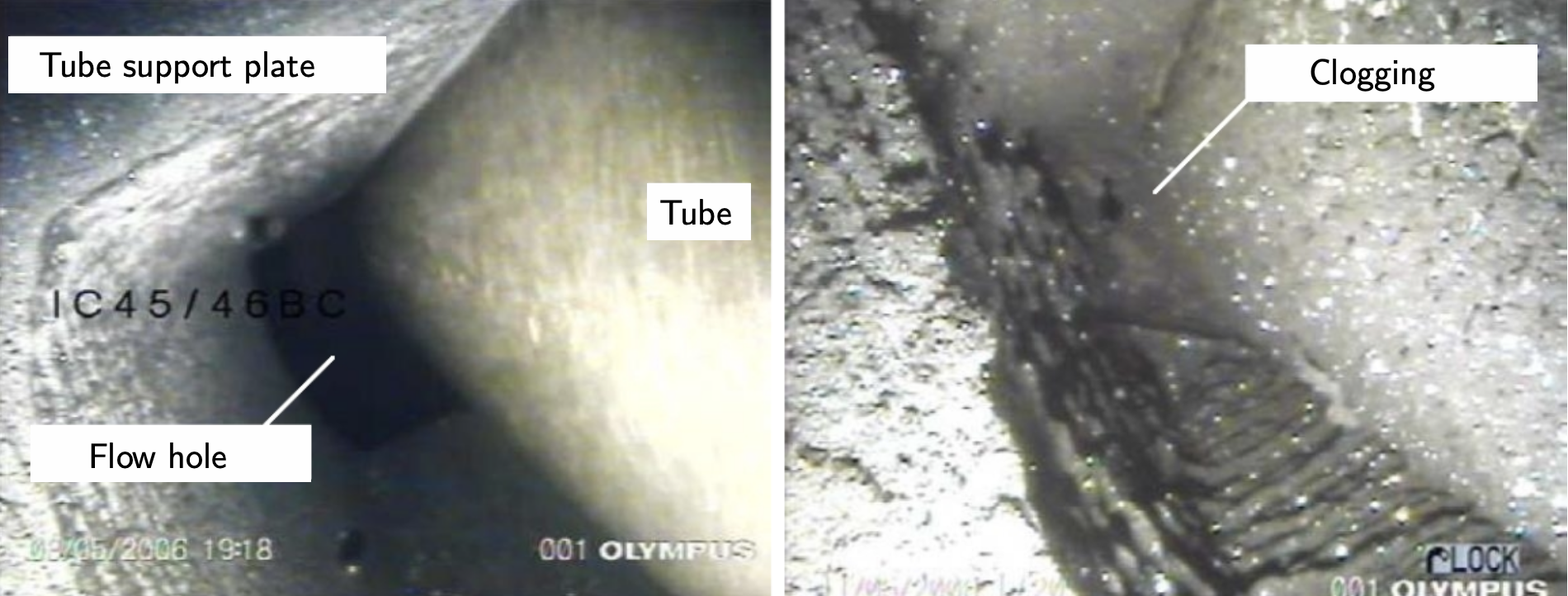} 
    \end{minipage}
\caption{PWR scheme, and example of video examination during an PWR outage (\textcopyright~IRSN, EDF).}
\label{fig:pwr_sg_scheme}
\end{figure}

Steam generators (SGs) are essential components of pressurized water reactors (PWRs), functioning as heat exchangers between two distinct water circuits. In PWRs, water from the primary circuit is first heated in the reactor pressure vessel by nuclear reactions within the core. The heated primary fluid then flows into the SG, where it transfers energy to the secondary circuit. This setup typically includes three or four loops, each equipped with an SG. In the SG, the primary fluid circulates through a bundle of U-shaped tubes, stabilized by tube-support plates. Simultaneously, the secondary fluid flows around these tubes and through the flow holes of the tube-support plates, absorbing heat and vaporizing into steam. The resulting steam exits the SG through upper outlets and drives turbines to generate electricity. The primary fluid, now cooled, returns to the reactor pressure vessel to be reheated. The whole process is summed up in Figure \ref{fig:pwr_sg_scheme}. Over time, the secondary fluid may become contaminated with solid and soluble particles, which can deposit on the tube-support plates leading to clogging. This phenomenon can reduce the heat transfer efficiency of the SG, therefore losing actual power output of the PWR, it can cause local steam flux imbalances inducing vibrations of the structure that can lead, in extreme cases, to tube rupture. Therefore EDF periodically plans chemical cleaning maintenances to diminish the clogging rate and this operation is a significant cost for the company.

One of the primary challenges in addressing SG clogging is the scarcity of predictive models and controlled experiments. The existing literature on SG clogging is limited \citep{srikantiah2000, prusek2013, Girard2014, yang2017}, and current research has largely focused on diagnosing SG clogging using indirect measurements rather than developing predictive models. The system’s complexity, arising from intricate operational settings, diverse physico-chemical mechanisms, and sparse data, makes it difficult to model and places it within the domain of complex systems. A significant issue is the absence of controlled laboratory experiments for clogging. One of the few models capable of predicting SG clogging over long time scales — spanning the operational lifetime of a nuclear reactor — is the THYC-Puffer-DEPO (TPD) code \citep{Feng,jaber_et_al_2023}. This advanced multiphysics computational chain code generates clogging trajectories $(t\mapsto \tau_{c}(t))$ over simulation periods exceeding $60$ years by coupling multiple inner codes simulating different physical phenomena. The lack of experimental data complicates the validation of the existing physical model. Nevertheless, for effective maintenance planning and ensuring safety of the asset, the development of a reliable model to predict SG clogging is a crucial task for EDF. As we will see, our methodology is fully suited to tackle this issue and more generally for complex systems like nuclear reactors, where available empirical data may be scarce.

\subsubsection{Clogging model}
\label{subsubsection431}

In broad terms, the physical model of SG clogging proposed in \citep{prusek2013, Feng} and further investigated in \citep{jaber_et_al_2023} is structured as an interconnection of three temporal model levels, enabling the simulation of clogging kinetics $(t\mapsto \tau_{c}(t))$ over large time periods. For the first two stationary phases, conservation equations for thermohydraulic quantities are employed, complemented by phenomenological closure laws. Subsequently, for the transport phase, stationary quantities are utilized for the mass fractions of solid and soluble particles. Furthermore, these temporal evolutions occur within a specific chemical conditioning environment (governing the pH), which itself may evolve over time. More details about the clogging model can be found in \citep{jaber_et_al_2023}.
The code TPD is a multiphysics solver of the clogging system. The THYC software developed by EDF R\&D \citep{david_THYC_nureth9_1999} is a numerical finite-volume scheme using a porous-medium approach to provide stationary thermohydraulic quantities at a component scale. The Puffer code is used for computing the solubility maps of magnetite $\Gamma^{\text{max}}_{s}$ in the secondary fluid, knowing a certain chemical conditioning, and finally the DEPO module is an iterative solver of the deposit equations. 
The TPD simulation code $g$ takes as inputs a set of chemical cleaning events gathered in $c \in \mathbb{N}$, performed in the past operating history of the SG. These cleaning events divide the resulting trajectories into $c+1$ segments. This gives rise to independent assimilation windows as described in the general methodology in \ref{fig:offline_hybrid}. Modeling the efficiency of these cleaning procedures is the subject of ongoing topics of research and modeling. Consequently, we can consider treating each segment independently between two cleanings in the assimilation process.

\subsubsection{Clogging data}
\label{subsubsection433}
\noindent There are two primary sources of information for monitoring the clogging rate $\tau_{c}$. As mentioned, there are no controlled experiments for the parameters of the TPD model. Consequently, we must rely on operational data measured at different times consisting of televised exams (images) and indirect measurements. The data are considered as $q=2$ heterogeneous data groups, where both data types are acquired with different standard deviations. 

\subsubsection*{Field data - Televised inspections (TVIs)}
\noindent Televised camera inspections allow for the estimation of a local clogging rate in every flow hole. This is based on image analysis methodology detailed in \citep{Girard2014}. Until now, this is considered, from an industrial viewpoint, as the most reliable clogging measure. However, it is scarce since it can be only realized during reactor outage.

\subsubsection*{Regression data}
\noindent In addition to TVIs, an in-house statistical methodology, largely inspired from \citep{pinciroli2021}, has been proposed for clogging rate estimation. The method consists in exploiting periodical transient state tests and various operational data to predict the clogging rate. It extracts features from complex operational time-series and uses indirect measures based on Foucault currents detailed in \citep{prusek2013}, providing a reasonably good monitoring of the clogging rate. However, this methodology lacks explicability and can perform poorly on different SG types.

\subsubsection{Design of experiments and metamodels}

\label{subsubsection432}
\begin{table}[h!]
    \centering
    \begin{tabular}{|c|c|}
        \hline
        \textbf{Input variable} & \textbf{Distribution} \\
        \hline
        $\alpha$ &  $\mathcal{U}[100, 103]$ \\
        $\beta$ & $\mathcal{U}[0.02, 0.025]$ \\
        $\epsilon_{e}$ & $\mathcal{U}[0.2, 0.5]$ \\
        $ \epsilon_{c}$ & $\mathcal{U}[0.01, 0.3]$ \\
        $d_p$ & $\mathcal{U}[0.5, 10.0] \times 10^{-6}$ \\
        $\Gamma_{p}(0)$ & $\mathcal{U}[1.0, 8.0] \times 10^{-9}$ \\
        $a_v$ & $\mathcal{U}[0, 15]\times 10^{-4}$\\
        \hline
    \end{tabular}
    \caption{Probabilistic modeling of uncertain input variables.}
    \label{tab:probabilistic_modeling_inputs}
    \end{table}

\begin{figure}[h!]
    \centering
    \includegraphics[width=1.0\linewidth]{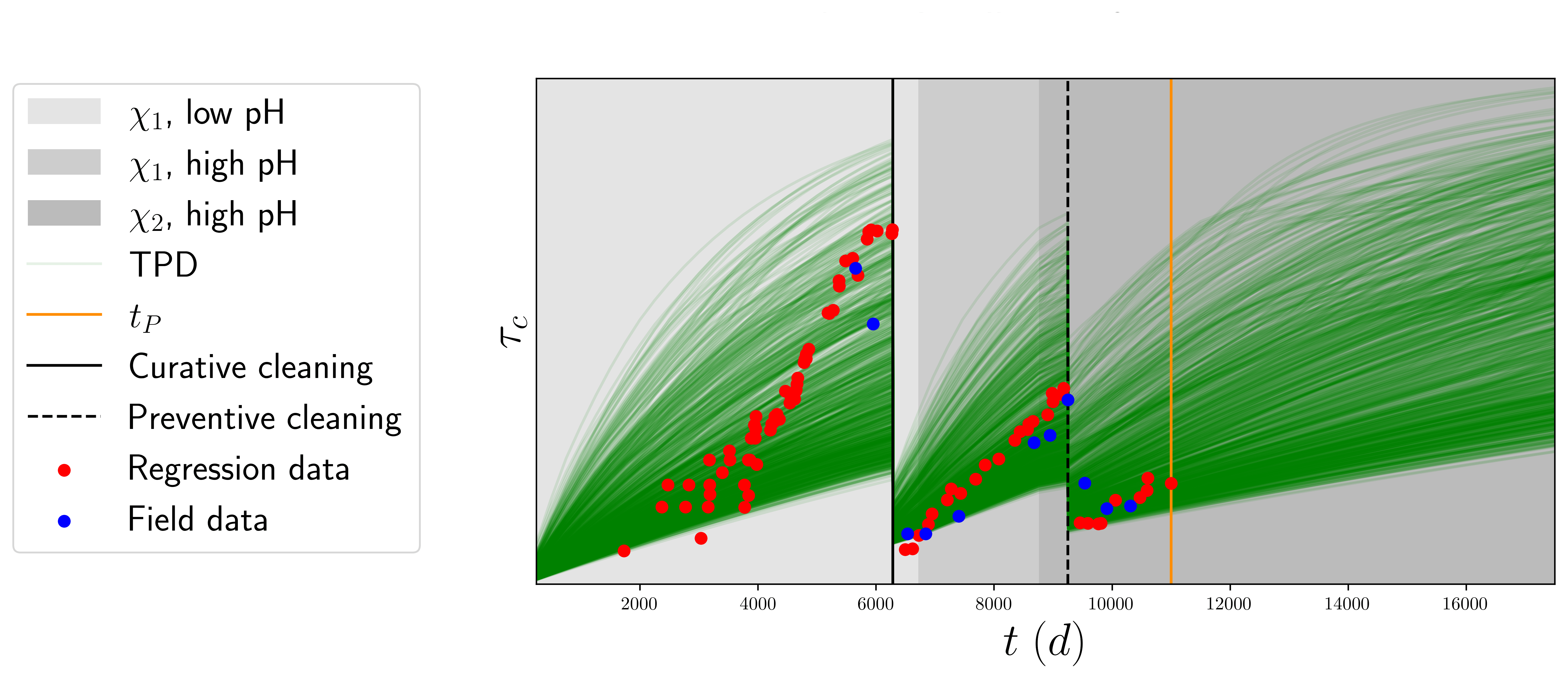}
    \caption{DoE of TPD simulations model for constructing different surrogates.}
    \label{fig5:designs}
\end{figure}

\noindent  The input variables and their distributions are summarized in Table~\ref{tab:probabilistic_modeling_inputs}. These parameters correspond to different physical parameters of the clogging model, more detail about them can be found in \citep{jaber_et_al_2023}. The TPD simulations are run on the EDF high-performance computing infrastructure, with one CPU node per simulation - a unitary call taking about $6$h for $60$ year operations of the SG. The first DoE amounts to $n=10^3$ simulations. A first full KLE metamodel is constructed on the output simulation time-grid using a constant mean and an absolute exponential kernel to predict the modes~\eqref{eq:kl_modes}. The resulting metamodel has a severe loss of predictive performance close to the discontinuities because of the chemical cleanings and otherwise a time-averaged $Q^2$ of about $0.9$. Afterwards, we train $p=12$ KLE metamodels on the linearly interpolated time instances of the data, with respective constant, linear and quadratic trends while considering both Matérn-$\nu$ with $\nu\in\{1/2, 3/2, 5/2\}$, squared exponential and absolute exponential covariance kernels. These metamodels are re-trained at each iteration step $k$ of the methodology described in Figure \ref{fig:general_scheme} on a new DoE following a sampling from the updated distribution $\mu_{\bm{X},k+1}$ where the prior distribution of $\theta_{k}$ has been replaced with the posterior.

\subsubsection{Numerical results}
\label{subsubsection434}
\begin{figure}[ht!]
    \centering
    \includegraphics[width=1.0\textwidth]{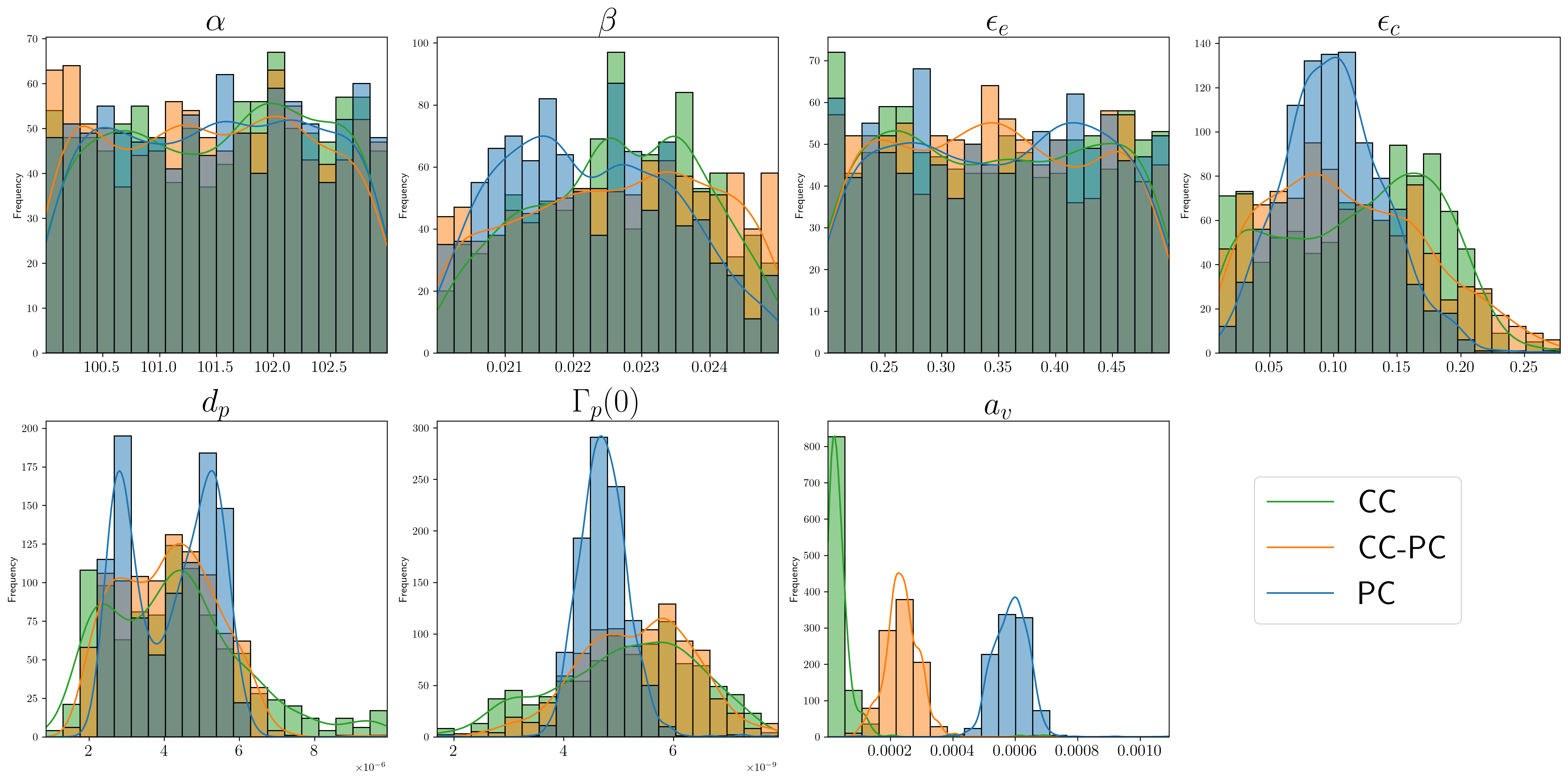}
    \caption{Posterior distributions of the input variables of TPD after the BMU step.}
    \label{fig:tpd_posterior_distributions}
\end{figure}

We perform $L=3$ independent assimilation procedures based on three time windows namely before curative cleaning (CC), between curative and preventive cleaning (CC-PC), and after the last preventive cleaning (PC) — corresponding to the current operational period of the SG. As can be seen from the results in Figure~\ref{fig:tpd_posterior_distributions}, the heterogeneous data fusion process has reduced the uncertainty in the input variables. In total, for each scenario, only $5$ input variables are selected in the sensitivity analysis step, namely $\beta, \epsilon_c, d_{p}, \Gamma_{p}(0)$ and $a_v$. At each iteration, $5$ MCMC chains are launched for computing the Gelman-Rubin statistic $R$ and convergence is assessed by checking if $R\simeq 1$. The full methodology on the three scenarios takes about $45$ min on a regular computer. For all the scenarios, the most influential variable at the first iteration is $a_{v}$, also called the \emph{vena contracta} calibration parameter \citep{prusek2013}, thereby confirming prior work done in \citep{jaber_et_al_2023}. The next iterations select parameters $\Gamma_{p}(0), d_{p}, \epsilon_{c}$ and the last considered is $\beta$. Each scenario ends with a re-selection and calibration of $a_{v}$ which does not change from the first iteration. 

For the sake of clarity, only the posterior distributions are displayed, as all the priors correspond to uniform distributions over the same support. The posterior distributions of parameters $\alpha, \beta, \epsilon_{e}$ remain nearly identical to their prior distributions, as shown in Figure~\ref{fig:tpd_posterior_distributions}, indicating that these parameters have minimal sensitivity to the output. In contrast, the distributions of input variables $\epsilon_{c}, d_{p}, \Gamma_{p}(0)$, and $a_{v}$ are significantly updated. This aligns with the sensitivity analysis results obtained in \citep{jaber_et_al_2023}. Notably, parameter $a_{v}$ exhibits three distinct modes corresponding to the different chemical cleanings. This behavior is linked to its role in the \emph{vena contracta} flux, where it acts as a linear parameter directly proportional to the clogging kinetics \citep{jaber_et_al_2023}. The observed decrease in deposition speed after each chemical maintenance is accurately captured by the three decreasing modes in the posterior distribution.

\begin{figure}[ht!]
    \centering
    \includegraphics[width=1.0\textwidth]{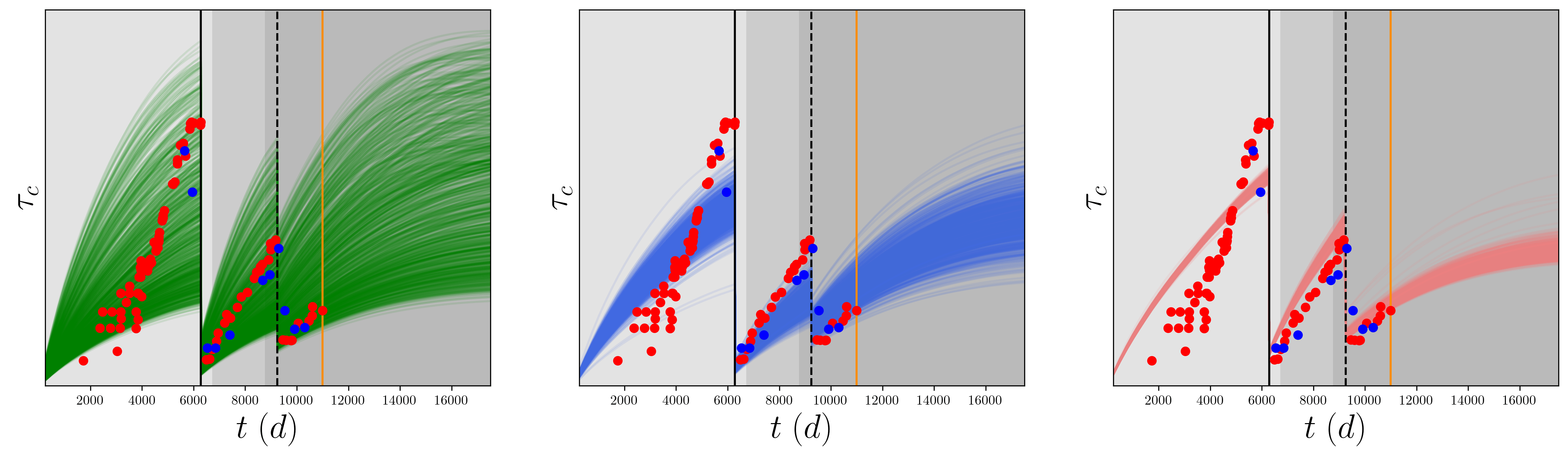}
    \caption{Trajectories of KLE-TPD emulator, from the prior distributions (left), after the BMU step (center) and after applying the Kalman smoothing step (right).}
    \label{fig:tpd_prior_posterior_trajs}
\end{figure}

The updated distributions of these influential parameters lead to a more precise prediction of the clogging trajectories. Figure~\ref{fig:tpd_prior_posterior_trajs} illustrates the prior and posterior trajectories of the clogging rate. The posterior trajectories are better concentrated around the observed data, demonstrating the effectiveness of the proposed data fusion methodology in reducing the input uncertainty. 

\begin{figure}[ht!]
    \centering
    \includegraphics[width=1.0\textwidth]{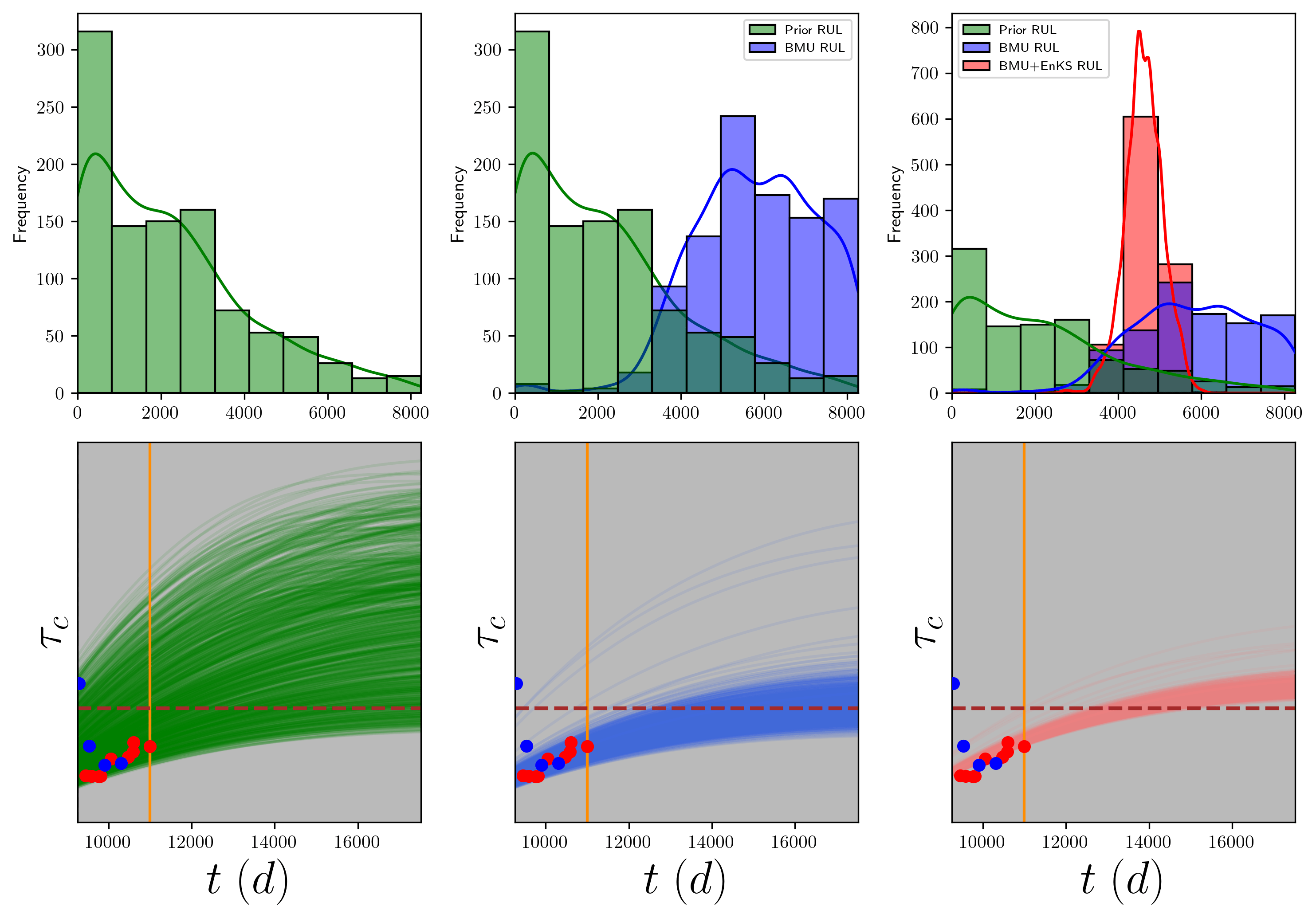}
    \caption{Different RUL distributions on the prognostics window, corresponding to the time-perior after the last preventive cleaning.}
    \label{fig:tpd_prior_posterior_ruls}
\end{figure}

Additionally, the methodology significantly improves the prediction of the RUL for a given threshold $D$. Figure~\ref{fig:tpd_prior_posterior_ruls} compares the prior and posterior RUL distributions after the last preventive maintenance, corresponding to the prognostics window in the methodology from Figure\ref{fig:offline_hybrid}. The posterior RUL distribution is narrower and more concentrated after both the BMU and the smoothing step, reflecting the reduced uncertainty and enhanced reliability of the predictions. This improvement is crucial for maintenance planning and operational decision-making, as it provides a more accurate estimate of the time remaining before the next maintenance is required.

\section{Conclusion and perspectives}
\label{sec5}
In summary we have presented a novel data fusion methodology for complex degradation phenomena of systems with scarce and heterogeneous data. The methodology is based on a Bayesian framework and can make use of Karhunèn-Loève expansions of output field functions as surrogates to update the prior distributions of the influential input variables when the dynamic simulation code is expensive to evaluate. The methodology is applied to two case studies: the first one is the Paris-Erdogan's law for crack growth prediction while the second one focuses on the clogging of steam generators in pressurized water reactors. The results show that the methodology is effective in reducing the uncertainty in the input variables and improving the precision of the RUL predictions, thus proving to be a valuable tool for assisting industrial decision making. Future work will focus on integrating the latent variables uncertainty within the MCMC procedure to obtain full posteriors, as well as on the development of adaptive metamodel updating strategy by quantifying uncertainty in the MCMC predictions by using, for instance, conformal prediction techniques \citep{leiWasserman2021,Jaber2025-2}.

\section{Acknowledgments}
\label{sec6}
The authors would like to thank Dr. Merlin Keller and Dr. Julien Pelamatti (EDF R\&D) for fruitful discussions about Bayesian calibration and field metamodeling with OpenTURNS. This work is part of a PhD program funded by the French National Association for Technological Research (ANRT) under Grant n\textsuperscript{o} 2022/1412.

\section{Appendix}
\label{sec7}
\subsection{Gaussian processes regression}
\label{subsec71}
\citep{rasmussen2006} Assume that $ g \sim \mathcal{GP}(m_{\beta}, \gamma_{\varphi}) $, where $ m_{\beta} : \Theta \to \mathbb{R}$ is the GP mean function expressed as an additive function $\beta_{\theta}^{\top} h(\theta)$. Parameters $ \beta^*_{\theta} $ are optimized along with the covariance kernel (e.g., Matérn type) $ \gamma_{\varphi} : \Theta^2 \to \mathbb{R} $, which also has hyperparameters $ \varphi^* $ to be optimized. The posterior mean is denoted by $ \widetilde{g}(\theta) $. Depending on the choice of prior kernel family, the number of hyperparameters may vary. For instance, for Matérn-$\nu$ kernels, $ \varphi = (\sigma, \ell)$ corresponds to the scale and correlation length parameters. Recall that Matérn-$\nu$ kernels are of the form:

\begin{equation}
K_{\varphi}(\theta, \theta') = \sigma \frac{2^{1-\nu}}{\Gamma(\nu)} \left( \sqrt{2\nu} \frac{\lVert \theta - \theta' \rVert}{\ell} \right)^{\nu} K_{\nu} \left( \sqrt{2\nu} \frac{\lVert \theta - \theta' \rVert}{\ell} \right),
\end{equation}
where $ K_{\nu} $ is a modified Bessel function of the second kind and $ \Gamma $ is the Euler gamma function. Note that this is an isotropic kernel. We also use exponential and Gaussian kernels, all of which are available in the OpenTURNS Python library (\url{http://openturns.github.io}). The posterior mean and kernel are defined for all $\theta, \theta' \in I$ as:
\begin{equation}
    \widetilde{g}(\theta):= k(\theta)^{\top}\bm{K}^{-1}g(\theta),
\end{equation}
and
\begin{equation}
    \widetilde{K}(\theta, \theta'):= K(\theta, \theta') - k(\theta)^{\top}\bm{K}^{-1}k(\theta'),
\end{equation}
where for all $\theta \in I$:
\begin{equation}
    k(\theta):=(K(\theta, \theta^{(1)}),\ldots,K(\theta, \theta^{(n)}))^{\top}\in\mathbb{R}^n,
\end{equation}
and
\begin{equation}
\bm{K}:= (K(\theta^{(i)}, \theta^{(j)}))_{1\leq i,j \leq n}\in\mathcal{M}_{n}(\mathbb{R}).
\end{equation}
To construct the TPD Gaussian process, we use the design of experiments $\text{DoE}^{*}_{\text{GP}} = \{\left(\bm{\Theta}, g(\bm{\Theta})\right)\}$.Hyperparameters $\varphi$ are optimized by minimizing the log-likelihood: 
\begin{equation}
    \varphi_{\text{MLE}} \in \argmin_{\varphi} \left\{g(\bm{\Theta})^{\top}\bm{K}^{-1}g(\bm{\Theta}) + \log(\det \bm{K}) \right\}.
\end{equation}

\subsection{Proof of Proposition~\ref{prop1}}
\label{subsec72}

\begin{proof}{}
    By Bayes' theorem:
    \begin{equation}
        p(\sigma^{2}_{\bm{\eta}}|\theta, \bm{y}) = \frac{p(\theta, \sigma^{2}_{\bm{\eta}}|\bm{y})}{p(\theta|\bm{y})} \Leftrightarrow p(\theta|\bm{y}) = \frac{p(\theta,\sigma^{2}_{\bm{\eta}}|\bm{y})}{p(\sigma^{2}_{\bm{\eta}}|\theta,\bm{y})}.
    \end{equation}
    Using equation~\eqref{eq:posterior_proba} and applying the assumptions, we obtain:
    \begin{equation}
        p(\theta,\sigma^{2}_{\bm{\eta}}|\bm{y})\propto p(\bm{y}|\theta)\lambda^{-1} = \lambda^{m/2 - 1}\exp\left(-\frac{\lambda}{2}\lVert \bm{y} - \bm{\mathcal{G}}(\theta)\rVert^{2}\right).
    \end{equation}
    Consequently:
    \begin{equation}
        p(\theta|\bm{y})\propto \frac{\lambda^{-1}p(\bm{y}|\theta)}{p(\lambda|\theta,\bm{y})} \propto \frac{\lambda^{m/2 -1}\exp\left(-\frac{\lambda}{2}\lVert \bm{y} - \bm{\mathcal{G}}(\theta)\rVert^{2}\right)}{\lambda^{m/2-1}\lVert \bm{y} - \bm{\mathcal{G}}(\theta)\rVert^{2m/2}\exp\left(-\frac{\lambda}{2}\lVert \bm{y} - \bm{\mathcal{G}}(\theta)\rVert^{2}\right)} = \lVert \bm{y} - \bm{\mathcal{G}}(\theta)\rVert^{-m}.
    \end{equation}
    Assuming the heterogeneous groups of data, we get by Bayes' theorem:
    \begin{equation}
        p(\theta|\bm{y}^{1},\ldots,\bm{y}^{q}|\bm{y}^{1},\ldots,\bm{y}^{q}) = \frac{p(\theta,\lambda_{1},\ldots,\lambda_{q})}{p(\lambda_{1},\ldots,\lambda_{q}|\theta,\bm{y}^{1},\ldots,\bm{y}^{q})}
    \end{equation}
    By writing the densities, we get
    \begin{equation}
        \begin{aligned}
       p(\theta|\bm{y}^{1},\ldots,\bm{y}^{q}) &\propto \frac{\prod_{i=1}^{q}\lambda_{i}^{m_{i}/2 - 1}\exp\left(-\frac{\lambda_{i}}{2}\lVert \bm{y}^{i} - \bm{\mathcal{G}}_{i}(\theta)\rVert^{2}\right)}{\prod_{i=1}^{q}\lambda_{i}^{m_{i}/2 - 1}\lVert \bm{y}^{i} - \bm{\mathcal{G}}_{i}(\theta)\rVert^{m_{i}}\exp\left(-\frac{\lambda_{i}}{2}\lVert \bm{y}^{i} - \bm{\mathcal{G}}_{i}(\theta)\rVert^{2}\right)}\\
       &\propto \prod_{i=1}^{q}\lVert \bm{y}^{i} - \bm{\mathcal{G}}_{i}(\theta)\rVert^{-m_{i}}.
        \end{aligned}
    \end{equation}
\end{proof}
\label{app1}

\bibliographystyle{elsarticle-num} 
\bibliography{hybrid}

\end{document}